\documentclass[%
 reprint,
 amsmath,amssymb,
 aps,
]{revtex4-2}
\usepackage{tikz}
\usepackage{braket}
\usepackage{lipsum}  \usepackage{booktabs}
\usepackage{graphicx}
\usepackage{dcolumn}
\usepackage{bm}
\usepackage[dvipsnames]{xcolor}
\usepackage{siunitx}
\usepackage{hyperref}

\usepackage{ulem}

\begin{document}

\title{Exact quantum scars from kinetic frustration for cross-platform realizations
}
\author{Zhuoli Ding$^{1}$}
\author{Ruben Verresen$^{2}$}
\author{Zoe Z.~Yan$^{1,\dagger}$}

\affiliation{$^1$ James Franck Institute and Department of Physics, The University of Chicago, Chicago, IL 60637, USA\\
$^2$ Pritzker School of Molecular Engineering, The University of Chicago, Chicago, IL 60637, USA\\
}

\date{\today}

\begin{abstract}
Quantum many-body scars are nonthermal states exhibiting persistent revivals in an otherwise ergodic, nonintegrable quantum system. Here we leverage the phenomenon of kinetic frustration -- the destructive interference of multiple quantum paths -- to create exact scars. The simplicity makes these models directly suitable for implementation on multiple existing quantum simulation platforms. In particular, we show how frustrated hardcore bosons in cold atom Bose-Hubbard simulators and polar molecule or Rydberg atom tweezer arrays have persistent oscillations whose lifetimes can be tuned with experimentally accessible parameters, like the Hubbard interaction or a Floquet drive. Second, we propose an experimentally realizable scar within a non-integrable Fermi-Hubbard model where the frustration arises from the fermionic exchange statistics, which admits a one-to-one mapping with the bosonic model in the scar subspace. Finally, we introduce a practical heuristic based on the energy distribution of eigenstates for systematically predicting and optimizing quantum many-body scar lifetimes.
Their cross-platform realizability and long lifetimes make them well-suited for benchmarking coherence and exploring nonergodic dynamics in current and near-term quantum devices.

\end{abstract}

\maketitle

\textit{Introduction:}
Quantum frustration plays a central role in strongly correlated matter, where it enhances quantum fluctuations and can stabilize exotic phases such as spin liquids and unconventional superconductors~\cite{fradkin2013field}. One particularly transparent form is kinetic frustration, in which destructive interference between competing hopping paths constrains particle motion. The advent of highly coherent quantum simulators has opened up new ways of exploring the consequences of kinetic frustration in and near-equilibrium~\cite{struck2011quantum, Leung2020, semeghini2021probing, chen2023continuous, Mongkolkiattichai2023, xu2023frustration, prichard2024directly, qiao2025kinetically, geim2026engineering, bornet2026dirac}. However, its consequences for far-from-equilibrium many-body dynamics remain much less understood, whereas this is precisely a regime of strongly interacting quantum systems that quantum simulators can unlock.

Here we show that kinetic frustration provides a simple route to experimentally accessible nonergodic many-body dynamics. For selected initial states, destructive interference can suppress ergodic spreading and confines the evolution to a special sector of Hilbert space, leading to persistent oscillations. Remarkably, this can arise even at infinite temperature in nonintegrable systems. Such revivals are a hallmark of quantum many-body scars~\cite{turnerWeakErgodicityBreaking2018, Serbyn2021QMBSReview, chandranQuantumManyBodyScars2023}, first discovered experimentally in constrained PXP-type models~\cite{bernien2017probing} and since explored both in theory~\cite{Turner18a,Choi19,Shiraishi2017ETHCounterexamples, Mori2017ThermalizationWithoutETH, Ok2019TopologicalScars, Kuno2020FlatBandScar, Omiya2023RydbergScars,Mark2020Scars, hudomal2020quantum, Moudgalya2020EtaPairing, ODea2020Scars, Pakrouski2020GroupInvariant, Ren2021Quasisymmetry,feng2025uncovering,Moudgalya2018AKLT, Schecter2019Spin1XY,moudgalya2022quantum, Russomanno2022, dong2023disorder, yang2025constructing,Moudgalya24,Lin19,Mark20,Pakrouski21,Langlett22,Desaules21_proposal,Kunimi24, gupta2026exact} and experimentally
in lattice models with cold atoms~\cite{bernien2017probing, bluvstein2021controlling,Su2023,Zhao2025} and superconducting circuits~\cite{Chen2022,dong2023disorder, zhang2023many}.
Our work identifies a minimal physical mechanism for realizing and probing such scarred dynamics, while also enabling tunable and parametrically long lifetimes.

We demonstrate this mechanism in minimal two-leg ladders with two complementary realizations of kinetic frustration. In the bosonic case, hopping in a $\pi$-flux background induces destructive interference between competing hopping paths, yielding an exact scarred subspace and perfect revivals from simple product states. In the fermionic case, the same interference structure emerges from exchange statistics, giving rise to scarred dynamics in a non-integrable Fermi-Hubbard-type model. In fact, the two descriptions are in one-to-one correspondence within the scarred sector. These models are also directly amenable for exploration in
at least two distinct types of quantum hardware platforms relevant to today's technologies: hardcore atomic bosons in $\pi$-flux optical lattices~\cite{Cooper2019} and dipolar spin-1/2 chains of Rydberg atoms~\cite{browaeys2020many} or polar molecules~\cite{langen2024quantum, cornish2024quantum}. 
We present numerical simulations demonstrating persistent measurable oscillations in these platforms and study experimental tuning knobs such as Floquet driving to enhance scar lifetimes.
Our model is therefore readily accessible in state-of-the-art experiments, an intriguing prospect given the scarcity of observations of tuneable quantum many-body scars.
Finally, we observe a general relationship between scar lifetimes and the energy distribution of eigenstates, giving an easily accessible heuristic that we use to characterize scar lifetimes.\\
\begin{figure}[t]
\includegraphics[width=0.48\textwidth]{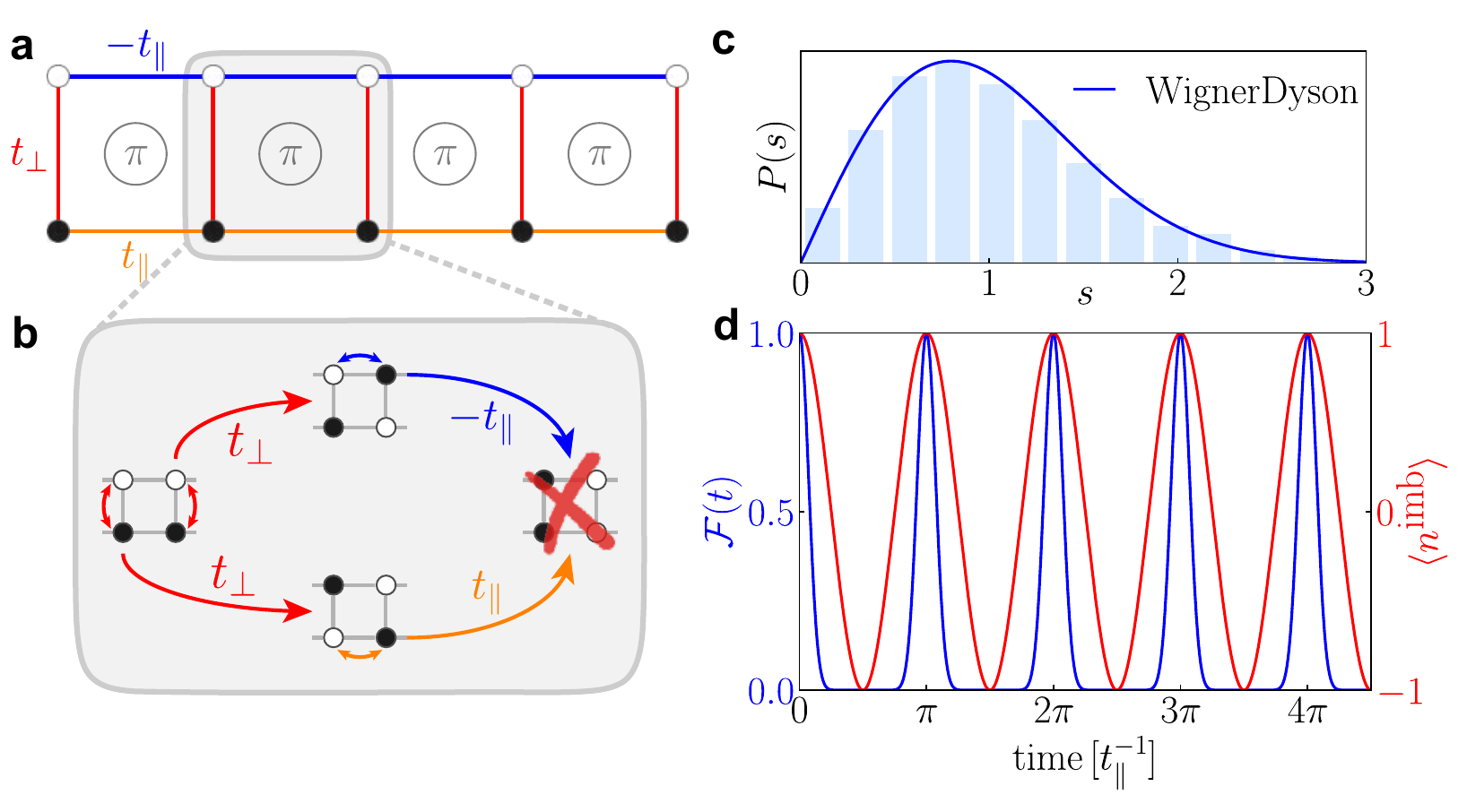}
\caption{\label{fig:1} 
\textbf{Exact quantum many-body scars in a hardcore boson ladder with $\pi$-flux per plaquette. (a)} Illustration of a $L = 5$ ladder with hardcore bosons tunneling as $-t_\parallel$ in the top leg, $t_\parallel$ in the bottom leg, and $t_\perp$ on the rungs. Our QMBS state is pictured as a filled bottom leg (black circles) and an unoccupied top leg. The destructive interference from tunneling around one square plaquette leads to a $\pi$-flux. 
\textbf{(b)} In one plaquette, destructive interference of the initial state (left) evolving into a final, non-scar configuration (right) is illustrated, via two potential paths (top and bottom) whose amplitudes carry a relative $\pi$ phase shift.  
\textbf{(c)}
The nearest-neighbor energy level spacing distribution of the model defined in Eq.~\ref{eq:exactscar} for a $L=9$ ladder with a small second-nearest neighbor hopping that preserves the scar (see End Matter for details). The spacings follow a Wigner-Dyson distribution (blue curve), indicating non-integrability of the model.
\textbf{(d)} Dynamics of the scar, showing the time-evolution of the many-body fidelity $F(t)$ (left axis, blue line) and the observable $\langle n^{\textrm{imb}}\rangle$ (right axis, red line) for a $L = 10$ ladder with $t_\perp = t_\parallel$.
}
\end{figure}
\textit{~Scars from kinetic frustration in bosonic platforms.}
The construction of the QMBS is motivated by a simple picture of kinetic frustration on a square plaquette.  
Consider a square in which hardcore bosons hop on four outside bonds with amplitude $|t|$, but the top bond has a hopping sign opposite to all others ($-t$).
These hardcore bosons are fully frustrated: there is destructive interference around a plaquette. 
We note that by the Peierls substitution method~\cite{Hofstadter1976}, this is equivalent to charges hopping in a magnetic field with $\pi$-flux on the plaquette. 
To see the effect for two bosons,
let us consider the time-evolution from a special initial state 
$|\psi\rangle=$~{\tiny $\left| \begin{matrix} 0 & 0 \\ 1 & 1 \end{matrix} \right\rangle$} with all bosons on the bottom leg.
Time-evolving does not allow for two bosons to occupy the same vertical rung, like {\tiny $\left| \begin{matrix} 0 & 1 \\ 0 & 1 \end{matrix} \right\rangle$}. Indeed, the $\pi$-flux implies destructive interference that cancels out the two pathways (see Fig.~\ref{fig:1}b). 

This motivates the definition of our QMBS, extending the system beyond a simple square plaquette to a $2\times L$ ladder, as the product state with all bosons in the bottom leg $|\psi_{\rm scar}\rangle=~{\tiny \left| \begin{matrix} 0 & 0 &0 \\ 1 & 1 &1 \end{matrix} ~~...\right\rangle}$. 
We denote the bosons as $a_j$ ($b_j$) on the top (bottom) leg. The Hamiltonian is (see Fig.~\ref{fig:1}a):
\begin{align} \label{eq:exactscar}
H &= -t_\perp \sum_j^L \left( a_j^\dagger b_j^{\vphantom \dagger} + {\rm h.c.}\right) 
+ t_\parallel \sum_j^L \left( a_j^\dagger a_{j+1}^{\vphantom \dagger}  + {\rm h.c.}\right) \\
&- t_\parallel \sum_j^L \left( b_j^\dagger b_{j+1}^{\vphantom \dagger}  +  {\rm h.c.}\right), \notag
\end{align}
where the hopping $t_\parallel$ along the legs and $t_\perp$ along the rungs can always be taken to be non-negative by absorbing a phase into the bosons. 
There are $N$ total number of sites, where $L=N/2$ is the length of the ladder. 
In the End Matter, we prove that the aforementioned kinetic frustration is sufficient for $|\psi_{\rm scar}\rangle$ to exhibit persistent oscillations of the particles on each rung.
We note that a scar family in a similar model (with additional disorder in the on-site potential) was studied theoretically and experimentally with superconducting circuits in Ref.~\onlinecite{dong2023disorder}, which demonstrated the existence of ergodicity-breaking initial states with anomalously low entanglement entropy. However, this was in the context of rainbow scars~\cite{Langlett22} instead of frustration; we highlight that kinetic frustration provides a simple but powerful mechanism, which we later also generalize to the fermionic case. Here we show the model of Eq.~\ref{eq:exactscar} is quantum chaotic and hosts an exact scar amenable to experimental implementation, which we furthermore numerically demonstrate through direct time-evolution for relevant experimental parameters. While frustration-based scars have appeared in projector-embedded models~\cite{McClarty20,Lee20}, our mechanism relies instead on kinetic frustration without invoking frustration-free degeneracies or flat bands~\cite{Creutz}.

\textit{Diagnostics of the exact scar.} 
A common figure of merit for the QMBS is the many-body fidelity of the initial state, $\mathcal{F}(t)\equiv |\langle \psi_{\rm scar}(t=0) | \psi_{\rm scar}(t)\rangle |^2$, but this observable is difficult to measure experimentally. 
Instead, we define another figure-of-merit: the average imbalance 
$\langle n^{\mathrm{imb}} \rangle = \frac{1}{L}\sum_{j=1}^L n^{\textrm{imb}}_j$, with $n^{\textrm{imb}}_j \equiv n^{\rm top}_j - n^{\rm bot}_j$, which is accessible to measurement in existing quantum simulators.
We numerically solve the model by performing exact diagonalization using QuSpin~\cite{10.21468/SciPostPhys.2.1.003}. We first show the distribution of energy level spacings in a symmetry sector (Fig.~\ref{fig:1}c), which follows the Wigner-Dyson distribution rather than Poissonian, indicating nonintegrability of the model where we added a small second-nearest neighbor hopping to minimize finite-size effects (see End Matter). 
Fig.~\ref{fig:1}d shows the time evolution of $\mathcal{F}(t)$ and averaged imbalance $\langle n^{\textrm{imb}} \rangle$, where the exact revival is confirmed by full-contrast many-body oscillations that persist forever. 
This simple, disorder-free model finds wide application across several hardware platforms, of which we highlight two: (1) hardcore atomic bosons in optical lattices in the presence of artificial magnetic fields, and (2) spin-1/2 particles interacting under the dipolar XY (spin-exchange) Hamiltonian.

\begin{figure*}[t]
    \includegraphics[width=0.9\textwidth]{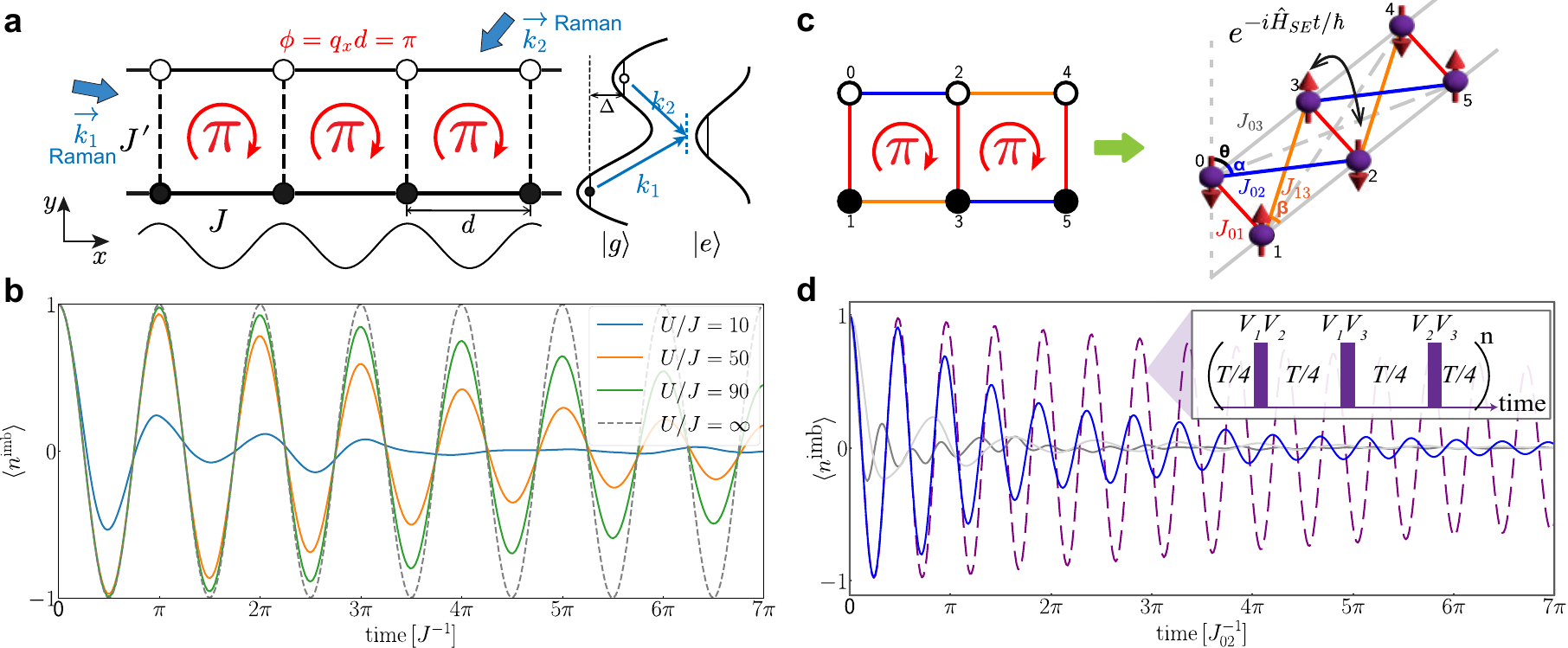} 
\caption{\label{fig:2} \textbf{Experimental implementations of QMBS in the bosonic model. (a)} In the implementation with ultracold atoms under an artificial magnetic field, the optical lattice is created by a uniform sinusoidal potential in $x$ and a staggered double-well in $y$, with energy offset $\Delta$ and lattice constant $d$.  A pair of Raman lasers (blue arrows) with wavevectors $\mathbf{k_1,k_2}$ induce photon-assisted tunneling in $y$ via an optical two-photon transition using the atom's excited state $|g\rangle\rightarrow|e\rangle$.  Nearest neighbor complex- and real-valued tunnelings with magnitudes $J',J$, respectively, couple the $y$ and $x$ directions. 
With appropriate choice of Raman laser detunings and angles, each plaquette has an effective $\pi$-flux. 
\textbf{(b)} Dynamics of the above scar for $L$ = 7, showing $\langle n^{\textrm{imb}} \rangle $ for various $U/|J|=U/|J'|$ in different colors. For infinite interactions, the bosons obey a hardcore constraint, leading to persistent revivals.  
\textbf{(c)} Implementation of the model using dipolar spin-1/2 degrees of freedom by transformation of the two-leg hardcore boson ladder of Fig.~\ref{fig:1}a into a dipolar zig-zag chain.
Particles trapped in optical tweezers form a spin-1/2 system spanned by $\ket{\downarrow}$ and $ \ket{\uparrow}$, and they interact under the spin-exchange Hamiltonian $\hat{H}_{SE}$.
Relevant parameters of the chain with six sites are pictured. Intra-rung interactions (in red) $J_{01}$ couple $n,n+1$ with $n$ even.
Next-nearest-neighbor couplings (in solid grey) are zero at the angle $1-3\cos^2\theta=0$.  Blue and orange lines indicate $J_{02}$ and $J_{13}$, with angles of $\alpha$ and $\beta$, respectively.
Longer-range couplings are indicated in dashed grey lines.
\textbf{(d)} Dynamics of the dipolar QMBS showing $\langle n^{\rm imb}\rangle$ (in blue).  The scar lifetime can be extended by Floquet-engineering (in purple) to cancel the longer-range couplings, with the Floquet pulse sequence given in the inset (see text for details). 
The long-lived scar exists in contrast to typical thermal states (in grey), which show fast decay of oscillations (see End Matter for details).
}
\end{figure*}

\textit{Hardcore bosons in $\pi$-flux lattices.} 
Ultracold atoms in optical lattices have become a paradigm for analog quantum simulation~\cite{gross2017quantum}.
Bosonic atoms are commonly used in state-of-the-art quantum gas microscopes~\cite{gross2021quantum}, where site-resolved population dynamics are probed in high-fidelity snapshot images.
Here, we provide a concrete roadmap to realize our model using spinless bosons in an optical lattice within the Bose-Hubbard regime.
The protocol is inspired by proposals~\cite{jaksch2003creation,Mueller2004, gerbier2010gauge} and past experiments~\cite{Aidelsburger2013, Miyake2013, atala2014observation, tai2017microscopy, leonard2023realization, impertro2025strongly} 
for creating effective magnetic fields with Raman lasers.
The geometry we consider is pictured in Fig.~\ref{fig:2}a: a ladder in which bosons hop along one leg (in the $x$-axis) in a sinusoidal potential with periodicity $d$, and between legs (in $y$) within a `tilted' double-well system. 
The two legs experience an energy offset $\Delta$ much larger than the nearest-neighbor hoppings $t$, which are equal for inter- and intra-leg tunneling.
Therefore, inter-leg hopping is normally suppressed; however, a pair of Raman lasers with wavevectors $\mathbf{k}_1,\mathbf{k}_2$ ($|\mathbf{k}_2|\approx |\mathbf{k_1}|$) can provide photo-assisted tunneling between legs~\cite{Aidelsburger2013,Miyake2013}. 
Experimentally, a Hamiltonian similar to Eq.~\ref{eq:exactscar} is realized when the Raman lasers experience a frequency difference $\hbar(\omega_1-\omega_2) = \Delta$, creating a running wave along the vector $\mathbf{q} = \mathbf{k_1-k_2}$.
This results in spatially-dependent complex tunneling amplitudes $J$ and $J' = |J'|e^{i\phi(\mathbf{r})}$ along $x$ and $y$, respectively, leading to an Aharonov-Bohm phase $\phi= q_xd$ when a particle tunnels around a plaquette, which can be tuned to $\phi=\pi$. 
Doubly-occupied sites experience a Hubbard repulsion, $U$, which blocks multiparticle occupancies.
The effective model governing the system under the tight-binding approximation is the Harper-Hofstadter-Bose-Hubbard (HHBH) Hamiltonian~\cite{Harper1955, Hofstadter1976}  [see Supplement].
These experimental components have been demonstrated on cold atom simulators~\cite{Aidelsburger2013, Miyake2013, atala2014observation, tai2017microscopy, leonard2023realization, impertro2025strongly}. 

The QMBS is realized by initializing $L$ bosons on one leg of the $2\times L$ ladder (Fig.~\ref{fig:2}a) -- a product state that was already demonstrated in past experiments~\cite{impertro2025strongly}. 
The QMBS manifests as the population dynamically changing from all bosons in the bottom leg to all bosons in the top leg.  
This oscillation repeats indefinitely for a perfect scar, as shown in the dashed line of Fig.~\ref{fig:2}b. 
The hardcore boson constraint is approached by working in a regime where $U/|J'|, U/|J| \gg 1$.
At weaker Hubbard repulsion, the initial state will leak into states in the Hilbert space with double occupancies, causing decay of the scar observable. 
Using Feshbach resonances for tuning $U$ will give access to the relevant strongly-interacting regimes capable of probing the QMBS with long-lived revivals.

\textit{Dipolar spin chains.} 
Beyond hardcore bosons in optical lattices, the QMBS can be implemented in a dipolar spin-1/2 zig-zag chain (Fig.~\ref{fig:2}c), with a boson mapped to $\ket\uparrow$ and an empty site mapped to $\ket\downarrow$. 
The spins can be encoded in two rotational levels of a polar molecule or two orbital angular momentum states of a Rydberg atom [see Supplement], and motion of the particles is frozen.
The system evolves under the lattice spin-exchange Hamiltonian~\cite{Barnett2006,gorshkov2011,Hazzard2013}$\hat{H}_{\rm SE} = \sum_{i>j} \frac{J_{ij}}{2}\left(\hat{S}_i^+\hat{S}_j^-+\hat{S}_i^-\hat{S}_j^+\right)$, where $S^+,S^-$ are the spin-1/2 raising and lowering operators.
The exchange interaction is given by the dipolar coupling  $ J_{ij} =V_{\rm ddi}(\mathbf{r_i-r_j}) = \frac{d^2(1-3\cos^2\theta)}{4\pi\epsilon_0 |\mathbf{r_i-r_j}|^3}$, where $\theta$ denotes the angle between the dipoles' quantization axis and their separation and $d$ is the electric dipole moment.
Instead of a ladder geometry, we ``twist" every other rung to obtain the zig-zag chain of Fig.~\ref{fig:2}c, and the tunnel-couplings of Eq.~\ref{eq:exactscar} become the exchange interactions $J_{ij}$. 
The carefully chosen angles in the chain are crucial to create the QMBS. 
By the choice of the zig-zag angles $\alpha,\beta$ and the quantization axis direction (which can be out of plane), the diagonal coupling terms $J_{\rm 13}$ and $J_{\rm 02}$ are equal in magnitude but opposite in sign, fulfilling the   kinetic frustration condition of destructive interference in analogy to the model in Fig.~\ref{fig:1}a.  
Crucially, the undesired interactions $J_{03}, J_{12}$ 
can be avoided using the node in the coupling at $\theta=\cos^{-1}{\sqrt{1/3}}$ [see Supplement for the full set of parameters]. 

The QMBS state is prepared in an initial product state with $L$ sites excited to $\ket{\uparrow}$ in an $2L$-length chain:
$\ket{\psi_{\rm scar}}=\ket{\downarrow\uparrow\downarrow\uparrow\downarrow...}$ with $\langle n^{\textrm{imb}} \rangle=1$.
Fig.~\ref{fig:2}d shows the evolution of $\langle n^{\textrm{imb}}\rangle$ for a zig-zag $L=10$ chain with $\alpha = 34^\circ, \beta = 32.45^\circ$ and the quantization axis $30^\circ$ out of the plane. 
Compared to a typical thermal state, $\ket{\psi_{\rm scar}}$ demonstrates long-lived emergent many-body oscillations. 
Unlike the exact scar of the Hamiltonian in Eq.~\ref{eq:exactscar}, here, the oscillations decay over time due to the presence of longer-range dipolar couplings (shown as grey dashed lines in Fig.~\ref{fig:2}c).

To further stabilize the scar dynamics, we use simple Floquet engineering~\cite{bluvstein2021controlling, Hudomal} to cancel the longer-range (grey dashed) interactions, which are the dominant errors in the dipolar chain. 
Considering a $\pi$ rotation on rung $j\in [1,L]$, we define $V_j {\equiv} \exp\big(\textrm{i}\pi(a_j^\dagger a_j{+}b_j^\dagger b_j)\big)$ in the hardcore boson ladder notation. Let $U(t) = \exp(-\textrm{i} t \hat{H}_{\rm SE})$ denote the time evolution. 
We define the Floquet unitary over a period $T$, defined for a unit cell of four rungs:
$ \label{eq:floquet_1}
U_F(T) = U(\frac{T}{4})V_1V_2U(\frac{T}{4})V_1V_3U(\frac{T}{4})V_2V_3U(\frac{T}{4})
$.
This Floquet operator cancels out the longer-range couplings over four rungs that reduce the scar lifetime, while preserving the nearest-neighbor couplings $J_{\rm 01}$ and the relationship $J_{\rm 02}=-J_{\rm 13}$, when averaged over many periods $T$ [see End Matter]. 
Fig.~\ref{fig:2}d shows the scar observable with and without Floquet driving, demonstrating that the cancellation of longer-range couplings leads to enhanced stability of the QMBS. One can similarly systematically cancel higher-order terms if desired.

Our proposal can be implemented in existing Rydberg and polar molecule quantum simulators.
In dipolar Rydberg chains~\cite{browaeys2020many, Barredo2015, de2019observation, chen2023continuous, bornet2023scalable, chen2025spectroscopy, Emperauger2025}, nearest-neighbor couplings of $J\approx h\times 0.1-10$MHz are achievable.
Decoherence primarily stems from finite lifetimes and spreading of position distributions due to thermal motion.
An estimate of these effects using values reported in Ref.~\cite{chen2025spectroscopy} show that the scar visibility is robust against this leading disorder effect for certain parameter regimes [see Supplement].  
In polar molecules, the spin-exchange Hamiltonian has been studied in lattice~\cite{yan2013observation, christakis2023probing, li2023tunable,carroll2025observation} and optical tweezer~\cite{holland2023demand, bao2023dipolar, picard2025entanglement,ruttley2025long} systems, and extension to our model with a dipolar zig-zag chain is straightforward.

\begin{figure}[htb]
    \centering
    \includegraphics[width=0.95\linewidth]{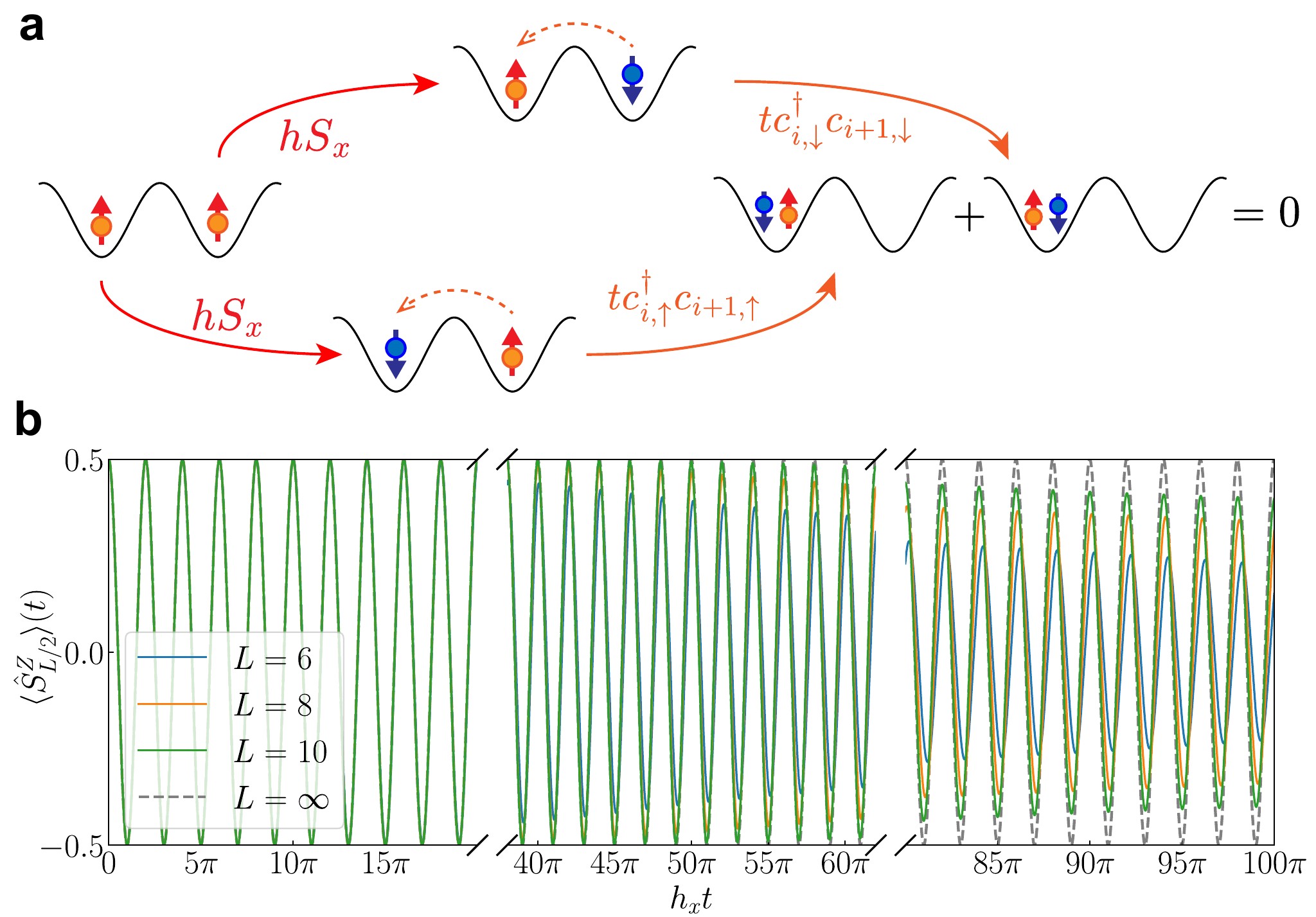}
    \caption{\textbf{QMBS in a generalized Fermi-Hubbard model} \textbf{(a)} For neighboring sites, destructive interference of the initial state (left) evolving into a final, non-scar configuration (right) is illustrated, via two potential paths (top and bottom) whose amplitudes cancel out due to the fermionic statistics. \textbf{(b)} Dynamics in a 1D chain of fermions in open boundary conditions for different system sizes, with  magnetic field $H_B = \sum_i \left(h_z \hat{S_j}^Z + h_x \hat{S_j}^X\right)$, where $h_z, h_x$ are the relative strengths of magnetic fields along $z$ and $x$. 
    $\langle\hat{S}^Z_{L/2}\rangle$ is the expectation value of $\hat{S}^Z$ for the central site. Simulation details can be found in the End Matter. The scar shows perfect revival when $L = \infty$ or with PBC (gray dashed curve) while slowly decay in OBC (blue, yellow and green curve). The decay speed becomes slower as system size increases.}
    \label{fig:3}
\end{figure}

\textit{Exact scars in a generalized Fermi-Hubbard model.}
Beyond scars in bosonic $\pi$-flux ladders, kinetic frustration can be used to construct QMBS in other systems, which we illustrate with a generalized Fermi-Hubbard model. Let us consider the time-evolution of a 1D spin-1/2 fermionic chain from a special initial state with unit filling of spin-up fermions with magnetic field along $x$. Two pathways that lead to double occupancy of fermions cancel out with each other due to the fermionic statistics, as shown in Fig~\ref{fig:3}a. This mechanism can be connected to the bosonic case of Eq.\ref{eq:exactscar} via a Jordan-Wigner transformation [see End Matter], and leads to persistent many-body oscillations between all spin-up to all spin-down fermions. 

However, these ``scar" dynamics can also be explained by the spin rotational SU(2) symmetry at zero magnetic field~\cite{Moudgalya2020EtaPairing} with the $x$-field leading to global spin precession. To test our mechanism, we introduce an additional $W$ term that, together with the $x$-field, breaks all continuous spin symmetry while preserving the kinetic frustration and hence the scar: $H_W = \sum_{i\neq j}W_{ij}\left( \hat{n}_i \hat{S}_j^Z + \hat{n}_j \hat{S}_i^Z \right)$, where $\hat{c}_{i\sigma}^\dagger$ is the creation operator of a fermion at site $i$ with spin $\sigma$, $\hat{n}_i = \sum_\sigma \hat{c}_{i\sigma}^\dagger \hat{c}_{i\sigma}$ and $\hat{S}_i^Z = \frac{1}{2}\left(\hat{c}_{i\uparrow}^\dagger \hat{c}_{i\uparrow} - \hat{c}_{i\downarrow}^\dagger \hat{c}_{i\downarrow}\right)$. 
The QMBS initial state is $\ket{\hat{S}^Z = +1/2}^{\otimes N}$. We numerically verified the presence of exact QMBS in 1D and 2D with periodic boundary conditions in Fig~\ref{fig:3}b (see Supplement).
This Hamiltonian can be implemented with lattice-trapped, fermionic polar molecules experimentally, where the dipole-dipole interactions naturally produce $H_W$~\cite{gorshkov2011, carroll2025observation} (see Supplement).

\begin{figure}[t]
\centering
    \begin{minipage}{0.49\textwidth}
        \centering
        \includegraphics[width=\textwidth]{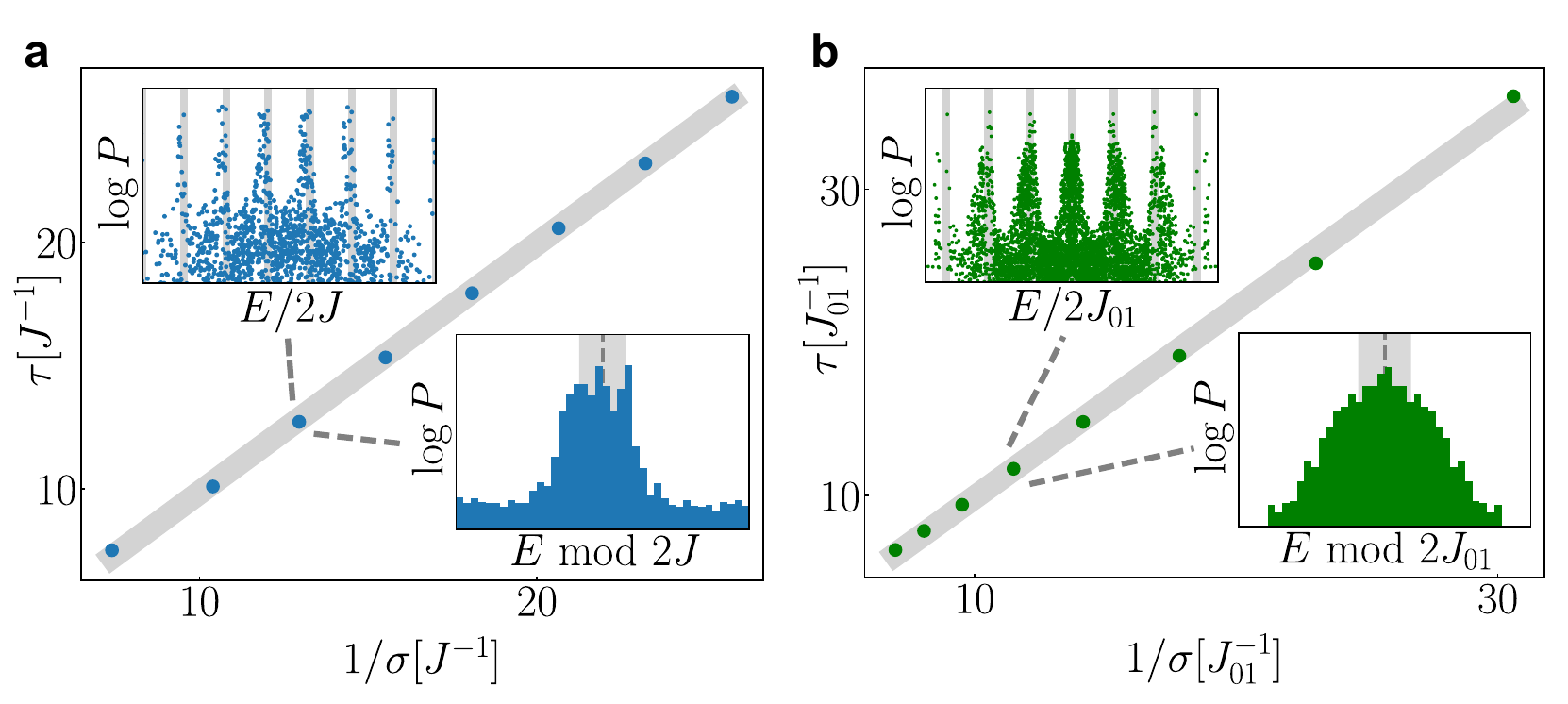}
    \end{minipage}
\caption{\label{fig:4}  \textbf{Predicting the thermalization time of an approximate QMBS from its energy distribution.} 
Due to experimental limitations, the approximate QMBS states exhibit finite lifetimes $\tau$ in both (\textbf{a}) the $\pi$-flux boson ladder taking $|J'|=|J|$ and (\textbf{b}) the dipolar spin chain, where each point is a different parameter -- varying the Hubbard $U$ or the angles for the dipolar zig-zag chain, respectively [see Supplement for details].  
The top left insets show the eigenenergy distribution for one such Hamiltonian, revealing a spread compared to an exact scar's equally spaced tower of eigenstates (grey lines).
From this, we extract the fractional energy distribution (bottom right insets), fitting the width $2\sigma$ given by the grey shaded area. Here $P$ is the weighted probability distribution of the overlap between the QMBS state and many-body eigenstates.
There is a clear linear relationship between $\tau$ and $1/\sigma$ for both models, allowing us to optimize the scar lifetime using the energy distributions. 
}
\end{figure}

\textit{ Connection between scar lifetimes and energy distributions:} In the experimental realizations, imperfections (such as finite Hubbard interaction or the longer-range dipolar interactions) lead to a decay of the scar dynamics. 
In searching for parameters that maximize scar lifetimes, we find a useful connection between the lifetime and the energy distribution of the eigenstates. The periodicity $\tau = \frac{\pi}{ t_\perp}$ as in Fig.~\ref{fig:1} is a result of the scar being constructed out of a tower of energy eigenstates~\cite{Moudgalya2018AKLT, Schecter2019Spin1XY,Moudgalya2020EtaPairing} with spacing $\Delta E=\frac{2\pi}{\tau}=2t_{\perp}$ (i.e., $2|J'|$ in the Bose-Hubbard ladder and $2J_{01}$ in the dipolar spin chain) [see Supplement for details on the spectrum-generating algebra].
With imperfections, the spectral weight of the approximate QMBS state remains concentrated near these equally spaced energies [see Fig.~\ref{fig:4}], but the spread of this distribution leads to decay of the revivals. 
To quantify this, we first compute the overlap of the approximate QMBS state with the many-body eigenstates. 
For each contributing eigenstate, we define the \textit{fractional energy} as $E\bmod \Delta E$ -- the minimal distance to the nearest eigenenergy of the exact scar spectrum.
The inset of Fig.~\ref{fig:4} shows the resulting probability distribution, and the energy width $\sigma$ is its standard deviation.
We find that this energy width is inversely proportional to the lifetime of the approximate QMBS state [Fig.~\ref{fig:4}].
We extract the approximate scar's lifetime $\tau$ for a variety of experimental parameters by fitting the scar evolution to an envelope function $\sim\exp(-t/\tau)$ [see Supplement].
The linear relationship between $\tau$ and $1/\sigma$ suggests a strategy for optimizing scar dynamics. 
By directly computing the spectral width associated with a given initial state and Hamiltonian, one can assess the robustness of scarring without resorting to long-time dynamical simulations, and we expect that this easily calculable heuristic could be used widely across different scarred systems. We used this fitting-free approach to study how the scar lifetime depends on the perturbations away from the ideal scar, observing $\tau$ to scale with the inverse perturbation strength for these two examples [see Supplement for further results]. 

\textit{Conclusion and outlook:} We have demonstrated that kinetic frustration is sufficient to give rise to an exact scar, as realized in a hardcore boson model and generalized Fermi-Hubbard model that are experimentally accessible to current quantum simulators.
Though experimental limitations cause leakage out of the scar subspace and eventual thermalization, we show that tuning experimental parameters -- the Hubbard repulsion or Floquet engineering -- can bring the system parametrically close to the ideal exact scar. 
We have developed a heuristic for optimizing scar lifetimes, which we expect is broadly applicable to other scarred systems, and our observation that the scar lifetime scales with the inverse perturbation strength deserves further study in light of other parametric observations in the literature~\cite{Lin20}.
Our proposal could lead to the implementation of long-lived QMBS states, which could serve as a useful resource for storing long-lived quantum information~\cite{Serbyn2021QMBSReview}, generating entangled states for quantum metrology~\cite{Dooley2022}, and benchmarking coherence in quantum simulators across multiple platforms.

We would like to thank Zlatko Papic, Thomas Iadecola, Olexei Motrunich, Maksym Serbyn and Michael Zaletel for interesting discussions.
Z.Y. acknowledges support from the David and Lucile Packard Foundation (2024-77404), the Air Force Young Investigator Program (FA9550-25-1-0360), and the Neubauer Family Assistant Professors Program. 

\noindent $^{\dagger}$ Email: zzyan@uchicago.edu\\

\section*{End Matter}

\subsection{Exact Scar from Kinetic Frustration}

In Fig.~\ref{fig:1}b, we showed how the kinetic frustration of Eq.~\eqref{eq:exactscar} forbids two particles from hopping onto the same rung. Here we give a more detailed version of this argument to prove that $|\psi_{\rm scar}\rangle=~{\tiny \left| \begin{matrix} 0 & 0 &0 \\ 1 & 1 &1 \end{matrix} ~~...\right\rangle}$ is an exact scar which has infinitely long-lived oscillations where each particle moves up and down its associated rung, without ever delocalizing in the horizontal direction.

More generally, let $\ket{\phi}$ denote \textit{any} quantum state with the following two properties:
\begin{itemize}
    \item[(i)] each rung hosts exactly one particle; and
    \item[(ii)] the state is invariant under bond-centered left-right spatial inversion with eigenvalue $+1$. 
\end{itemize}
(Note that $|\psi_{\rm scar}\rangle$ satisfies these properties.)~We claim that time-evolution under our $\pi$-flux boson ladder preserves these two properties \textit{at all times}. More precisely, if we denote our model as $H(t_\perp,t_\parallel)$, then we now show that kinetic frustration implies $H(t_\perp,t_\parallel) \ket \phi = H(t_\perp,0) \ket \phi$. Thus, the effective dynamics is purely rung-dynamics, justifying our claim that $|\psi_{\rm scar}\rangle$ has persistent oscillations on each rung.

\textit{Proof:} A complete basis for each rung is given by the four states  $\ket{d^\pm}=~ \frac{1}{\sqrt{2}}(${\tiny $\left| \begin{matrix} 1  \\ 0  \end{matrix} \right\rangle~$}$\pm$ {\tiny $\left| \begin{matrix} 0  \\ 1  \end{matrix} \right\rangle$}), $\ket{d^0}= ~{\tiny \left| \begin{matrix} 0\\ 0 \end{matrix}\right\rangle}$ and $\ket{d^1}= ~{\tiny \left| \begin{matrix} 1\\ 1 \end{matrix}\right\rangle}$. By virtue of property (i), we know that in this basis, $\ket \phi$ only has overlap with $\ket{d^\pm}$ on each rung. By virtue of property (ii), each square plaquette only has overlap with the following three states:
\begin{equation}
\ket{d^+,d^+}, \quad \ket{d^-,d^-}, \quad \frac{\ket{d^+,d^-} 
+ \ket{d^-,d^+}}{\sqrt{2}}. \label{eq:threestates}
\end{equation}
Now we simply observe that kinetic frustration implies that the $t_\parallel$ term annihilates all these three states, by the reasoning in Fig.~\ref{fig:1}b. This means we stay within the states \eqref{eq:threestates}, which implies properties (i) and (ii). This proves our above claim.

\subsection{Floquet Engineering for the Dipolar Model}

While Floquet engineering has been used to improve scar lifetimes \cite{bluvstein2021controlling,Hudomal}, the mechanism is often non-trivial. In contrast, the knowledge of the ideal scar in our exact model allows for a straightforward pulse sequence that systematically cancels out the longer-range terms which would otherwise spoil the exactness. For instance, consider the pulse sequence from the main text:
\begin{align}
\footnotesize U_F(T) &= U\left(\frac{T}{4}\right)V_1V_2U \left(\frac{T}{4} \right)V_1V_3U \left(\frac{T}{4} \right)V_2V_3U \left(\frac{T}{4} \right) \notag \\
&= U\left(\frac{T}{4} \right)\left(V_{1} V_{2} \; U\left(\frac{T}{4} \right) \; V_{2}^\dagger V_1^\dagger\right) \\
& \quad \times \left( V_{2} V_{3} \; U\left(\frac{T}{4} \right) \; V_3^\dagger  V_{2}^\dagger  \right) U\left(\frac{T}{4} \right)  \notag
\end{align}
where $V_j$ denotes a $\pi$-pulse on the $j$'th rung. A hopping term is flipped by conjugating with the $V_j$'s if and only if it overlaps with only \textit{one} site involved in the pulse. This way, we see that the rung hoppings $t_{\perp}$ all add up, whereas the horizontal hoppings $t_{\parallel}$ get partially suppressed by a factor of two. Crucially, the second-nearest-neighbor hopping signs along the horizontal direction are flipped half of the time, thereby completely canceling out in this Floquet sequence.

\subsection{Generalized Definition of Imbalance}
For generic product states, we generalize the definition of imbalance: for each individual site as $i$, define $\sigma^z_i = 2n_i-1$, the averaged imbalance is defined as 
\begin{equation}\label{eq:imbalance}
    \langle n_{\textrm{imb}} \rangle = \frac{1}{N}\sum_{i=0}^N \langle \sigma^z_i(t) \rangle \langle \sigma^z_i(0) \rangle,
\end{equation}
where N is the total number of sites.

\subsection{Jordan-Wigner Transformation}
Here we show the details of Jordan Wigner transformation between the bosonic scar and fermionic scar. In Fig.~\ref{fig:1}(\textbf{a}), relabel the site index as ${\left( \begin{matrix} 0 & 3 & 4 \\ 1 & 2 &5 \end{matrix}~~...\right)}$. Define $c_j^\dagger = (-1)^{\sum_{k=1}^{j-1}n_k}b_j^\dagger$, where $b_j^\dagger$ is the creation operator of hardcore boson at position $j$. It can be shown that $\{c_j^\dagger\}$ satisfies ferimonic statistics and $c_j^\dagger c_{j+1} = b_j^\dagger b_{j+1}$. Moreover, for the additional ``long range" terms like $J_{03}, J_{25}$, we can show that $c_j^\dagger c_{j+3} = b_j^\dagger b_{j+3}(-1)^{n_j+n_{j+1}+n_{j+2}} = -b_j^\dagger b_{j+3}(-1)^{N_{j}^s}$, where $N_j^s = \sum_{k=j}^{j+3}n_k$ is the total boson number on a square plaquette. In the scar subspace, $N_j^s = 2$ for all j, so the mapping becomes $c_j^\dagger c_{j+3} = -b_j^\dagger b_{j+3}$, which contains an additional minus sign. In this way, the hardcore bosonic ladder with $\pi$ flux can be naturally mapped to a spinless fermionic ladder without flux, where the structure of kinetic frustration is preserved.
Furthermore, the $2\times L$ ladder of spinless fermions can be mapped to a $1\times L $ spin-1/2 chain, where the top (bottom) leg maps to $\ket{\uparrow} ~(\ket{\downarrow})$, leading to the picture of Fig.~\ref{fig:3}a.

\subsection{Numerical Details}

\textit{Details for Fig.~\ref{fig:1}.} In Fig.\ref{fig:1}c, we compute the nearest-neighbor level spacing distribution within the symmetry-resolved sector, for a chain of $L = 9$ with open boundary conditions and $t_\perp = t_\parallel$. Spin-flip symmetry, $S_z = 0$ conservation, and spatial reversal symmetry are taken into account.

To resolve the nonlocal symmetry~\cite{dong2023disorder}, we introduce next-nearest-neighbor tunneling in the horizontal axis, with $t_{\textrm{nn}}$ in the top leg and $-t_{\textrm{nn}}$ in the bottom leg. These additional terms preserve the conditions required for the existence of the scar state. The strength of the long-range hopping is chosen as $t_{\textrm{nn}} = 0.1 t_\perp$. After symmetry reduction, the effective Hilbert space dimension is 12120. The resulting level spacing statistics in the middle half of the energy spectrum yield an average gap ratio $r = 0.5291$, in close agreement with the Wigner-Dyson prediction $r_{\textrm{WD}} = 0.5295$. 
Further calculations indicating non-integrability of our model are presented in the supplement.

\textit{Details for Fig.~\ref{fig:2}.} For Fig.~\ref{fig:2}b, since the onsite interaction strength $U$ is much larger than the tunneling, we restrict the local bosonic occupation to $n_i \leq 2$ and choose $L = 7$ with open boundary conditions.

In Fig.~\ref{fig:2}d, we show the time evolution of the generalized imbalance of two representative product states. $|\psi_{\rm thermal,1}\rangle{=}{\tiny {\rm site ~index} \left| \begin{matrix} 0 & 3 & 4 \\ 1 & 2 &5 \end{matrix} ~~...\right\rangle} {=}{\tiny \left| \begin{matrix} 0 & 0 &0 \\ 1 & 1 &1 \end{matrix} ~~...\right\rangle}$ and $|\psi_{\rm thermal,2}\rangle{=}{\tiny {\rm  site ~index} \left| \begin{matrix} 0 & 3 & 4 \\ 1 & 2 &5 \end{matrix} ~~...\right\rangle} {=}{\tiny \left| \begin{matrix} 1& 0 &1 \\ 1 & 0 &1 \end{matrix} ~~...\right\rangle}$ correspond to grey and light-grey curves, respectively, using Eq.~\ref{eq:imbalance}.

\textit{Floquet engineering parameters in Fig.~\ref{fig:2}d}.  We used a Floquet period of $T = \pi/J_{01}$ and assumed infinitely short pulse durations $V_i$. We have separately confirmed that for pulse durations $\sim \frac{T}{20}$, the finite-pulse window has no appreciable effect on the scar quality.

\textit{Details for Fig.~\ref{fig:3}}. In Fig.~\ref{fig:3}b, we simulate the time dynamics of the generalized Fermi-Hubbard model with Hamiltonian: $H_{\textrm{FH}} = \sum_{\langle i,j \rangle, \sigma} t \left[ \hat{c}_{i\sigma}^\dagger \hat{c}_{j\sigma} + \text{h.c.} \right] + U \sum_i \hat{n}_{i\downarrow} \hat{n}_{i\uparrow} + \sum_{\langle i,j \rangle}W\left( \hat{n}_i \hat{S}_j^Z + \hat{n}_j \hat{S}_i^Z \right) + \sum_i \left(h_z \hat{S_j}^Z + h_x \hat{S_j}^X\right)$. We set $t = 1, U/t = 100, h_x = -h_z = 2, W = 1$ during the simulation (we stress that the scar exists for any value of $U/t$). For periodic boundary conditions, we simulate the time evolution in 1D with a $L$ = 10 chain and in 2D with a $3\times3$ lattice, which all show perfect oscillations. Discussions of the Fermi-Hubbard model in experiments and related parameters can be found in the supplement.

\textit{Details for Fig.~\ref{fig:4}}. In Fig.~\ref{fig:4}a, $\tau$ and $\sigma$ are calculated with L = 7 chain and $n_i \leq 2$, with $U/|J| = U/|J'|$ in the range of [30, 100]. In Fig.~\ref{fig:4}b, $\tau$ is fitted from the time evolution of a $L=10$ chain while $\sigma$ is calculated from exact diagonalization of a $L = 8$ chain. The quantization axis is $30^\circ$ out of the plane and $\alpha$ is varied in the range of [26, 40] degrees, corresponding to $J_{02}/J_{01}$ in the range of [0.2, 0.7]. The $y$ axis of top left inset ranges from [$10^{-7}$, 0.2] while the $y$ scale of bottom left inset is [$10^{-4}$, 1]. The fitting details can be found in the supplement. 

\bibliography{ref}

\setcounter{figure}{0}
\setcounter{equation}{0}
\setcounter{section}{0}

\clearpage
\onecolumngrid
\vspace{\columnsep}

\newcolumntype{Y}{>{\centering\arraybackslash}X}
\newcolumntype{Z}{>{\raggedleft\arraybackslash}X}

\newlength{\figwidth}
\setlength{\figwidth}{0.45\textwidth}

\renewcommand{\thefigure}{S\arabic{figure}}
\renewcommand{\theHfigure}{Supplement.\thefigure}
\renewcommand{\theequation}{S\arabic{equation}}
\renewcommand{\thesection}{\arabic{section}}

\begin{center}
    \large{\textbf{Supplemental Information: \\ }}
\end{center}

\section{Construction of the scar through the spectrum-generating algebra}

A useful way to define QMBS is through the explicit construction of a tower of special eigenstates using a ladder-operator~\cite{Moudgalya2018AKLT, Schecter2019Spin1XY,Moudgalya2020EtaPairing} which generates `ferromagnetic scar states' \cite{omiya2026generalized, o2026locality}.
On each rung, we define $\ket{d^\pm}=~ \frac{1}{\sqrt{2}}(${\tiny $\left| \begin{matrix} 1  \\ 0  \end{matrix} \right\rangle~$}$\pm$ {\tiny $\left| \begin{matrix} 0  \\ 1  \end{matrix} \right\rangle$}).
One can verify that the product state $\ket{\psi_0} = \ket{d^-d^-...d^-}$ is an eigenstate of $H$ due to the kinetic frustration. 

Following Ref.~\cite{chandranQuantumManyBodyScars2023}, we introduce the operators $J^\pm = \sum_{j=1}^{L}\ket{d^\pm_j}\bra{d^\mp_j}$ and $J^z = \sum_{j=1}^{L}(\ket{d^+_j}\bra{d^+_j}-\ket{d^-_j}\bra{d^-_j})$.
Repeated application of $J^+$ on $\ket{\psi_0}$ generates a sequence of eigenstates $\ket{\psi_n} \propto (J^+)^n\ket{\psi_0}$. The set of states generated in this way forms a subspace, which we refer to as the scarring subspace, $\mathcal{H}_{\text{scar}} = \text{span}\{\ket{\psi_0},\ket{\psi_1}...,\ket{\psi_L}\}$. Importantly, this subspace has dimension $L + 1$, and the eigenstates are equally spaced in energy by $\Delta E= 2t_{\perp}$. Consequently, any state within $\mathcal{H}_{\text{scar}}$ undergoes perfectly periodic time evolution and revives exactly after a fixed period of $\pi/t_\perp$. One can show that $\ket{\psi_{\mathrm{scar}}}$ can be written as a coherent superposition of all the scar eigenstates $\ket{\psi_n}$ and therefore this construction can provide an explicit explanation of the scarred nature of this special initial state.

In detail, the method of the spectrum generating algebra (SGA) states that for an eigenstate $\ket{\psi_0}$ with energy $E_0$, if the commutator between a creation operator $J^+$ and Hamiltonian $H_{\textrm{SG}}$ satisfies $[H_{\textrm{SG}}, J^+] = \omega J^+ (\omega \neq 0)$, then $\ket{\psi_n} = (J^+)^n \ket{\psi_0}$ is also an eigenstate with energy $E_0 + n\omega$ as long as the eigenstate is nonzero. We find that our model can be described by the notion of a restricted SGA~\cite{Moudgalya2018AKLT, Schecter2019Spin1XY,Moudgalya2020EtaPairing}, which only requires that the commutation relationship holds within the scarred subspace.  
The Hamiltonian can be written into three parts:
\begin{equation}
    H = H_{\textrm{sym}} + H_{\textrm{SG}} + H_{\textrm{A}},
\end{equation}
$H_{\textrm{sym}}$ commutes with $J^\pm$ and $J^z$, and $H_{\textrm{SG}}$ satisfies $[H_{\textrm{SG}}, J^\pm] = \pm \omega J^\pm$, lifting the degenerate multiplets of $H_{\textrm{sym}}$. The final term $H_\textrm{A}$ does not commute with $J^\pm, J^z$, but annihilates the scar states $H_\textrm{A}\ket{\psi_n} = 0$. Defining $\ket{d^0}= ~{\tiny \left| \begin{matrix} 0\\ 0 \end{matrix}\right\rangle}$, $\ket{d^1}= ~{\tiny \left| \begin{matrix} 1\\ 1 \end{matrix}\right\rangle}$, for the exact scar (Eq.~\ref{eq:exactscar}), $H_{\textrm{SG}}$ is 
\begin{equation}
    H_{\textrm{SG}} = -\sum_{j=1}^L t_{\perp}\bigg(\ket{d^+_j}\bra{d^+_j} - \ket{d^-_j}\bra{d^-_j}\bigg),
\end{equation}
$H_A$ is
\begin{equation}
\begin{split}
H_{A} &= t_\parallel \sum_{j=1}^{L-1}\bigg(
\ket{d^0_j d^{-}_{j+1}}\bra{d^+_j d^0_{j+1}} + \ket{d^-_j d^0_{j+1}}\bra{d^0_j d^+_{j+1}} \\
&- \ket{d^-_j d^1_{j+1}}\bra{d^1_j d^+_{j+1}} - \ket{d^1_j d^-_{j+1}}\bra{d^+_j d^1_{j+1}}
\bigg) + \mathrm{h.c.}
\end{split}
\end{equation}

We note that because the rung hopping ($H_{\rm SG}$) is not a single-site term, the persistent oscillations are also accompanied by nonzero entanglement oscillations, which physically corresponds to the particle on a given rung oscillating between the two sites as a periodic function of time. This is distinct from the mechanism discussed in Refs.~\onlinecite{o2025entanglement,o2026locality}, where in a variety of examples the energetics is determined by an effective \textit{single}-site Hamiltonian which thereby gives completely frozen entanglement dynamics.
We note this SGA, giving rise to a very similar scar family, was also discussed in the case of the ``rainbow scar" of Ref.~\cite{dong2023disorder}.

\section{Details of experimental implementations}\label{sec:1}
\subsection{Hardcore bosons in optical lattices: THE HHBH Hamiltonian}

Here we discuss the experimental implementation of the QMBS using hardcore bosons tunneling in an optical lattice under the Hofstadter-Bose-Hubbard (HHBH) Hamiltonian~\cite{Harper1955, Hofstadter1976}

\begin{equation} \label{eq:raman}
\begin{aligned}
\hat{H}_{\rm HHBH} &= -J' \sum_{m,n} \left( \hat{a}^\dagger_{m,n+1} \hat{a}_{m,n} e^{i \phi_{m,n}} + \text{h.c.} \right) \\
&\quad - J \sum_{m,n} \left( \hat{a}^\dagger_{m+1,n} \hat{a}_{m,n} + \text{h.c.} \right) \\
&\quad + \frac{U}{2}\sum_{m,n} \hat{n}_{m,n}(\hat{n}_{m,n}-1),
\end{aligned}
\end{equation}

Here, $J(J')$ is the effective hopping along $x(y)$, $a^\dagger_{m,n}~(a_{m,n})$ creates (annihilates) a boson on site $(x,y)=(m,n)$, $n_{m,n}$ is the number operator, and $U$ is the on-site Hubbard interaction.
Experimentally, the Hamiltonian in Eq.~\ref{eq:raman} is realized when the Raman lasers experience a frequency difference $\hbar(\omega_1-\omega_2) = \Delta$, creating a running wave along the vector $\mathbf{q} = \mathbf{k_1-k_2}$ [see main text Fig.~2].
When a particle tunnels around the boundary of one square plaquette, it picks up an Aharonov-Bohm phase.
With each site hosting a spatially dependent phase $\phi_{m,n} = \mathbf{q\cdot r_{m,n}}=q_x\,d\,m+q_y\,d\,n$, the effective flux is therefore $\Phi = q_x d$.
Appropriate choices of the angle between the Raman lasers and the frequency detuning can result in a uniform $\pi$-flux per plaquette in the ladder model.
The magnitudes of $J',J$ are functions of the bare hopping $t$, the phase difference between sites $\phi_{m',n}-\phi_{m,n}$, the Raman modulation amplitude, and the energy offset $\Delta$~\cite{aidelsburger2013experimental}. 
Importantly, the relative magnitudes of $J',J$ do not change the qualitative scar behavior, and all calculations are shown with $|J|=|J'|$.
We note that this protocol's main components have already been demonstrated on previous cold atom simulators~\cite{Aidelsburger2013, Miyake2013, atala2014observation, tai2017microscopy, leonard2023realization, impertro2025strongly}, though never in the context of quantum scars.
The requirement of a finite ladder with two legs can be achieved by isolating a $2 \times L$ subsystem out of an extended two-dimensional lattice using additional barrier laser fields shaped by spatial light modulators~\cite{tai2017microscopy}. 
Bosonic atomic species with tunable magnetic Feshbach resonances such as $^{7}$Li, $^{39/41}$K, and $^{133}$Cs could be used for implementing this scar.

\subsection{Dipolar spin-1/2 zig-zag chain}
The electric dipole-dipole interaction of two particles at $\mathbf{r_i,r_j}$ oriented by some external symmetry-breaking (e.g.~magnetic, electric) field is given by
\begin{equation} \label{eq:ddi}
    V_{\rm ddi}(\mathbf{r_i-r_j}) = \frac{d^2(1-3\cos^2\theta)}{4\pi\epsilon_0 |\mathbf{r_i-r_j}|^3}
\end{equation}
where $d$ is the electric dipole moment and $\theta$ parametrizes the angle between the quantization axis and the separation $\mathbf{r_i-r_j}$.
We consider a system where the dipoles' translational motion is frozen in a deep optical tweezer array.
Consider two internal states $\ket{\downarrow}, \ket{\uparrow}$ that represent opposite-parity states coupled by an electric dipole transition matrix element.  
In the absence of an external static electric field, the two-level system can be described with a lattice spin-exchange model~\cite{Barnett2006,gorshkov2011,Hazzard2013}
\begin{equation}\label{eq:spinexchange}
    \hat{H}_{\rm SE} = \sum_{i>j} \frac{J_{ij}}{2}\left(\hat{S}_i^+\hat{S}_j^-+\hat{S}_i^-\hat{S}_j^+\right)
\end{equation}
with the exchange interaction $J_{ij}$ given by the dipole-dipole interaction in Eq.~\ref{eq:ddi}, and $S^\pm_i$ are the spin-1/2 operators on site $i$. 
This Hamiltonian physically represents the transfer of the spin excitation between two particles; since a particle may not be doubly-excited, the spin excitation can be mapped to a hardcore boson.
The zig-zag chain with carefully chosen angles is a crucial ingredient to recover the phenomenology of the QMBS of Eq.~\ref{eq:exactscar}. Reconfigurable optical tweezer arrays allow for high tunability of couplings in an electric dipolar system.
By appropriate choice of quantization axis and the angles $\alpha,\beta$, we can ensure that the diagonal coupling terms $J_{\rm orange}$ and $J_{\rm blue}$ are equal in magnitude but opposite in sign (see Fig.~\ref{fig:3}c), fulfilling the kinetic frustration condition of destructive interference in analogy to the model in Fig.~\ref{fig:1}a.  

For state preparation of $\ket\psi_{\rm scar}$, we can start from a unit-filled array of $\ket{\downarrow}$.
The initial state can be prepared by selective excitation of the appropriate sites (1, 3, 5 ...) to $\ket{\uparrow}$ with microwaves, either in tweezers in a separate region which are then moved into position on the zig-zag chain, or directly in the target region along with site-selective dressing beams that provide local ac Stark shifts to isolate target sites for excitation.

Our proposal of the dipolar XY quantum many-body scar can be implemented in existing Rydberg and polar molecule quantum simulators.
Rydberg atoms trapped in optical tweezer arrays are a leading platform for studying quantum many-body dynamics~\cite{browaeys2020many}, and they can encode an effective spin-1/2 as opposite parity angular momentum states, \textit{e.g.}, $\ket{\uparrow}=\ket{nS}$ and $\ket{\downarrow}=\ket{nP}$, where $n$ is a high principal quantum number.
Thus they realize resonant dipolar interactions, which have already been utilized in several experiments~\cite{Barredo2015, de2019observation, chen2023continuous, bornet2023scalable, chen2025spectroscopy, Emperauger2025}.
The electric dipole moment $d$ in Eq.~\ref{eq:ddi} is given by the matrix element between $\ket{\uparrow},\ket{\downarrow}$, and an external magnetic field sets the quantization axis. 
The magnetic field can also isolate $\ket{\uparrow},\ket{\downarrow}$ from irrelevant Zeeman sublevels.
The resonant dipolar interactions lead to nearest-neighbor couplings $J\approx h\times 0.1-10$\,MHz at typical 10 - 20\,$\mu$m separations.
Decoherence primarily stems from finite Rydberg lifetimes ($\sim$100\,$\mu$s at room temperatures) and finite position distributions due to thermal motion ($\sigma_r\sim 100, \sigma_z\sim800\,$nm radially or axially in the tweezer)~\cite{chen2025spectroscopy}.
An estimate of these decoherence effects on the ideal approximate scar conditions show that the QMBS visibility still persists out to many oscillations (see Section \ref{sec:leakage}).  

In addition to Rydberg atoms, ultracold polar molecules are capable of realizing our model.  Polar molecules are emerging as a powerful quantum technology combining tunable interactions, rich internal structures, and long-lived coherence times~\cite{langen2024quantum, cornish2024quantum}.
Heteronuclear molecules possess permanent body-frame electric dipole moments of order 0.1-10 Debye, and their rotational levels $\ket{N,m_N}$ are strongly coupled through dipole-dipole interactions, where $N$ is the rotational quantum number and $m_N$ is its projection onto the quantization axis.
Within one electronic and vibrational level, molecules can form an isolated two-level system between $\ket{\uparrow}=\ket{N=0,m_N=0 }$ and any of $\ket{1,0}$ or $\ket{1,\pm1}$ for $\ket{\downarrow}$. 
The molecules are governed by the spin-exchange Hamiltonian of Eq.~\ref{eq:spinexchange}~\cite{micheli2006toolbox, Barnett2006, gorshkov2011}, as long as degeneracies between molecular nuclear spin states are lifted sufficiently to isolate the 2-level system (by nuclear quadrupole moments or by Zeeman shifts).
The spin-exchange Hamiltonian has been studied in lattice~\cite{yan2013observation, christakis2023probing, li2023tunable} and optical tweezer~\cite{holland2023demand, bao2023dipolar, picard2025entanglement,ruttley2025long} systems, and extension to our model with a dipolar zig-zag chain is straightforward.
Compared to Rydbergs, the nearest-neighbor $J$ of polar molecules is weaker (ranging from $h\times$10 Hz - 10 kHz levels at $\sim$ 1\,$\mu$m distances), but molecules benefit from long-lived coherence times of up to $\tau_2\approx10$\,s~\cite{ruttley2025long}, enabling decoherence-limited quality factors theoretically up to $J\tau_2\approx10^5$. 
However, state-preparation of large-scale defect-free molecular tweezer arrays is still an outstanding challenge.

\subsection{Fermionic Polar molecules in optical lattice: The t-J-V-W Hamiltonian}

Here we discuss the experimental implementation of the fermionic QMBS with polar molecules tunneling in an optical lattice. The related Hamiltonian is a generalized Fermi-Hubbard model~\cite{carroll2025observation},
\begin{align}
\hat{H} =& -h \sum_{\langle i,j \rangle, \sigma} t \left[ \hat{c}_{i\sigma}^\dagger \hat{c}_{j\sigma} + \text{h.c.} \right] + hU \sum_i \hat{n}_{i\downarrow} \hat{n}_{i\uparrow} + \notag \\ &h \sum_{i \neq j} \frac{V_{ij}}{2} [ J_\perp (\hat{S}_i \cdot \hat{S}_j) + \chi \left( \hat{S}_i^Z \hat{S}_j^Z \right) + V \hat{n}_i \hat{n}_j +   W \left( \hat{n}_i \hat{S}_j^Z + \hat{n}_j \hat{S}_i^Z \right) ]
\end{align}

Here $t$ is the effective hopping of nearest neighbor sites, $U$ is the onsite Hubbard interaction,  $\hat{c}_{i \sigma}^\dagger$ ($\hat{c}_{i \sigma}$) creates (annihilates) a fermion with spin $\sigma$ on site $i$, and $\hat{n}_{i \sigma}$ is the number operator. The final terms are from dipole-dipole interactions, where $V_{ij}$ is the geometric prefactor of the interaction between sites $i$ and $j$, $J_\perp$ is the Heisenberg exchange strength, $\chi$ is the exchange anisotropy, $V$ is a density-density interaction, and $W$ is a spin-density coupling. $J_\perp, \chi, V, W$ can be tuned by external electric or magnetic fields.
We note in the limit of high $U/t$, this becomes the $tJVW$ model where double occupancy is forbidden~\cite{gorshkov2011,carroll2025observation}.
Here we include $U$ both for generality as well as to make the connection with kinetic frustration (see Fig.~\ref{fig:3}). 

The Hamiltonian was first realized with fermionic KRb molecules in 3D optical lattices~\cite{carroll2025observation},where $U$ was neglected due to the low filling fraction of molecules. The existence of exact QMBS requires $\chi = 0$ and periodic boundary conditions. Nonzero $\chi$ term will break the internal structure of scar states. However, this term can be turned to 0 experimentally by the external electric field. For KRb molecules trapped in 1064nm optical lattice, $\chi = 0$ at $|E| = 6.5\textrm{kV/cm}$, and at this point, $J_\perp = 130\mathrm{Hz}, W = 23\mathrm{Hz}, V = 4\mathrm{Hz}$. The tunneling strength $t$ can be tuned by varying the lattice depth, with a range from 0 to 300\,Hz. The magnetic field along z is chosen to compensate $W$ term at unit filling.
Microscopically calculating Hubbard parameters for polar molecules is beyond the scope of this work, and we assume $U/t=100$ for the plots of Fig.~\ref{fig:3}. $U$ is tunable by choice of species, electric field strength, and lattice depth. We note that typically for unshielded molecules, chemical reactions lead to rapid two-body loss of doublons; it will be interesting to probe whether the QMBS can stabilize the system against such loss.

With open boundary conditions, sites in the boundary and in the bulk will experience different $W$ terms, which causes the scar to decay from the boundary. This effect is gradually suppressed as the system size increases.

\section{Level spacing distribution for periodic boundary conditions}

\begin{figure}[htbp]
    \centering
    \includegraphics[width=0.9\linewidth]{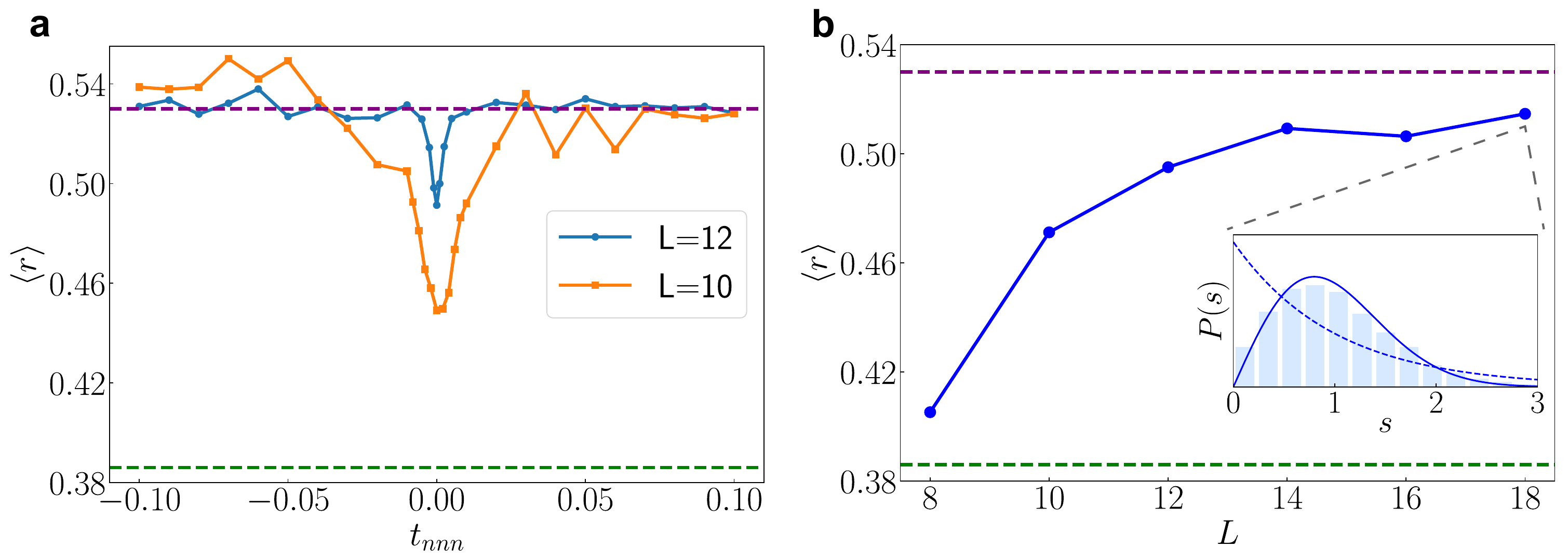}
    \caption{\textbf{Nearest neighbor spacing ratio with $t_{nnn}$ and $L$.} \textbf{(a)} $\langle r \rangle$ value in $S_z = 0, K_x = 0, P_y = 1, P_x' = 1, F = 1$ sector. $t_{nnn}$ is the next-next nearest neighbor hopping strength. The purple and green dashed lines indicate $\langle r\rangle$ for Wigner-Dyson and Poissonian distributions, respectively. \textbf{(b)} $\langle r \rangle$ value in $N_{\textrm{up}} = 6, K_x = 0, P_y = 1, P_x' = 1$ sector for different system size for $t_{nnn}=0$. The inset shows the level spacing distribution for $L = 18$. Blue and dashed blue curve in the inset correspond to Wigner-Dyson and Poisson distributions, respectively.}
    \label{fig:S0}
\end{figure}

Here we give additional details on the level spacing distribution, which indicates non-integrability of our model.
For the original model of Eq.~\ref{eq:exactscar} with periodic boundary conditions, label $P_y$ as the inversion along the $y$ axis, $F$ as the $Z$ flip symmetry, and $P_x'$ as inverting the top and bottom chain, and add $\pi$ phases to all the odd rungs. Next nearest neighbor interactions $t_{nn}$ introduced in the End Matter will break $P_x'$ so here we use the next-next neighbor interactions $t_{nnn}$ instead. $t_{nnn}$ preserves all symmetries and keeps the scar exact. We found that $\langle r \rangle$ has a dip compared to the value predicted by the Wigner-Dyson distribution at $t_{nnn} = 0$ (which is the model of Eq.~\ref{eq:exactscar}) and as $L$ grows, the dip becomes smaller, shown in Fig.~\ref{fig:S0}(a). To investigate whether the dip is due to the finite system effect, we also calculate $\langle r \rangle$ in $N_{\textrm{up}} = 6$ sector, where the Hilbert space dimension is greatly reduced and larger system sizes become accessible. In this sector, $\langle r \rangle$ increases with $L$ and at $L = 18$, $\langle r \rangle = 0.515$, which is close to Wigner-Dyson distribution, shown in Fig.~\ref{fig:S0}(b). 
Taken together, these data suggest that in the thermodynamic limit, our model with $t_{nnn}=0$ is non-integrable.

\section{Details about fitting in figure 4}
\begin{figure}[htb]
    \centering
    \includegraphics[width=0.95\linewidth]{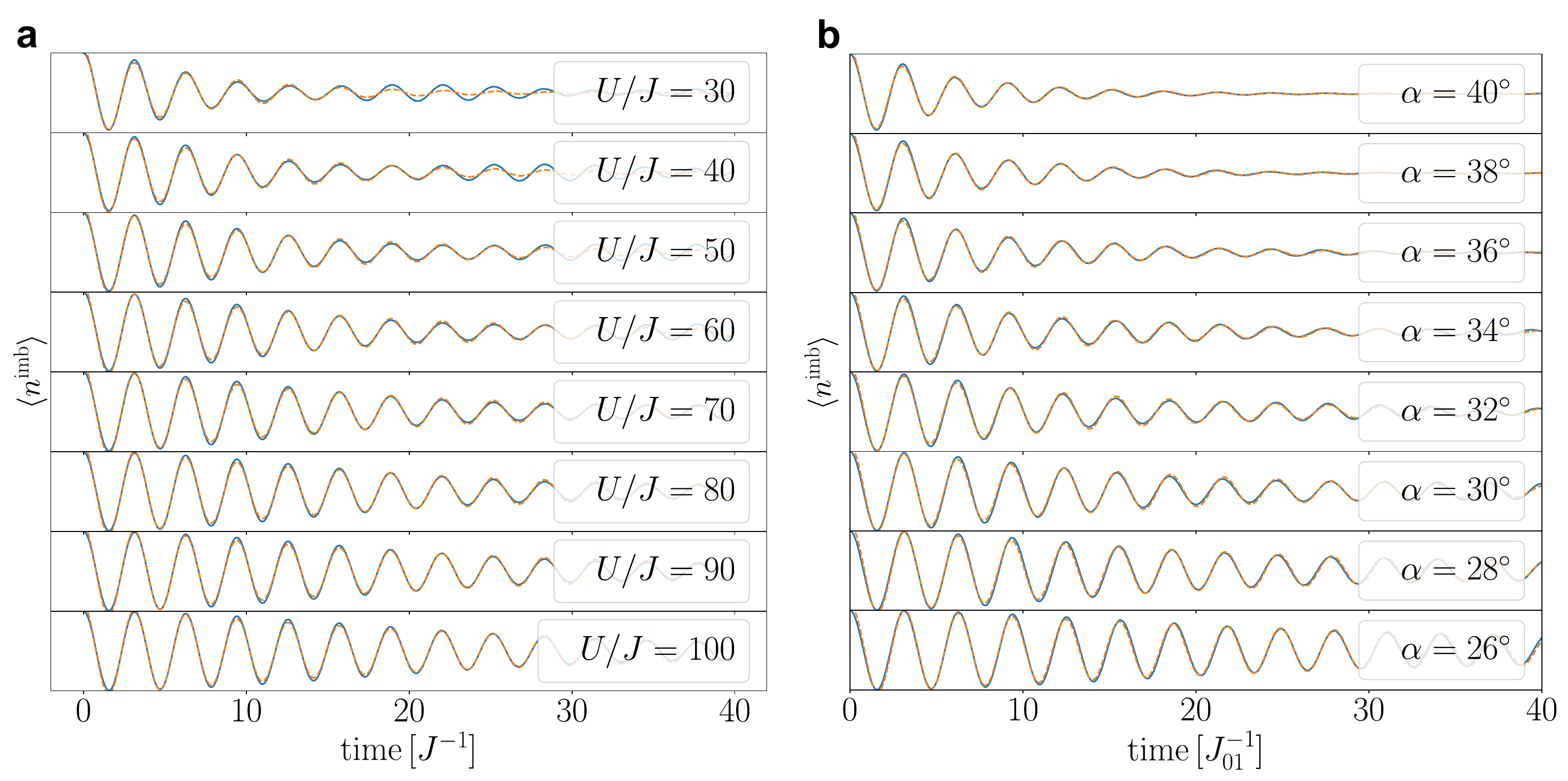}
    \caption{\textbf{Lifetime fitting details for Fig.~\ref{fig:4} of the main text.} Subfigure \textbf{(a)} shows the fitting results for optical lattice with different $U/|J|$ while \textbf{(b)} shows the results for dipolar spin chain under different $\alpha$. The blue curve is the simulated data from QuSpin and the yellow curve is the fit from Eq.~\ref{eq:lifetimefit}.
    \label{fig:S1}}
\end{figure}

The scar lifetime $\tau$ in Fig.~\ref{fig:4} is calculated by fitting the imbalance oscillation with time. The function used for fitting is 
\begin{equation}\label{eq:lifetimefit}
    f(t) = A\cos(\Omega t+\phi)\exp(-t/\tau)
\end{equation}
where A, $\Omega$, $\phi$ and $\tau$ are free fit parameters.
For bosonic lattice, the oscillation first decay and then revive, which is largely due to the finite system sizes. In order to minimize the finite system size effect, we choose the time that the oscillation amplitude reaches the first minimum as the end point of fitting. For dipolar spin chain, since the oscillation amplitudes never revive, we simply choose a cutoff at $J_{01}t = 40$. The fitting results for all the data points in Fig.~\ref{fig:4} of the main text are shown in Fig.~\ref{fig:S1}. The error bar of $\tau$ is smaller than the size of the point and thus is not shown.
\begin{figure}[ht]
    \centering
    \includegraphics[width=0.7\linewidth]{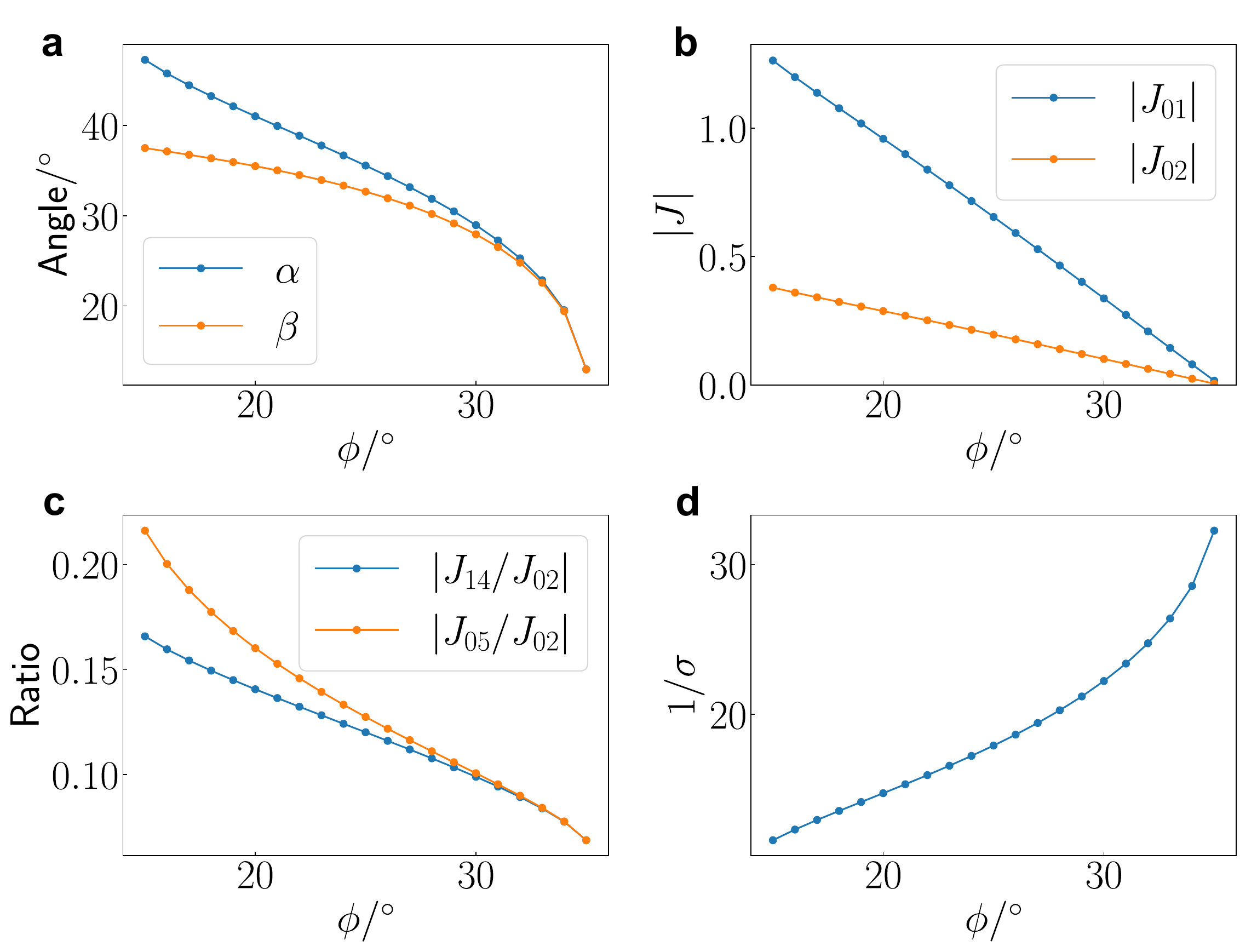}
    \caption{\textbf{Optimizing the scar lifetime for a dipolar spin chain.} Different parameters with respect to $\phi$ for a fixed $J_{02}/J_{01} = 0.3$. \textbf{(a)} $\alpha$ and $\beta$ values for realizing the model. \textbf{(b)} Long range interactions decrease with $\phi$. \textbf{(c)} Absolute value of the interaction strength decreases with $\phi$. \textbf{(d)} The estimated scar lifetime $1/\sigma$ with $\phi$ where $\sigma$ is calculated using the same method as in Fig.~\ref{fig:4}. }
    \label{fig:S2}
\end{figure}

\section{Finite system size effect}
Due to limited computational resources, the system size accessible to numerical simulations in QuSpin is limited to around $L = 10$, where boundary effects may have an influence on the result. 

\begin{figure}[htb]
    \centering
    \includegraphics[width=0.7\linewidth]{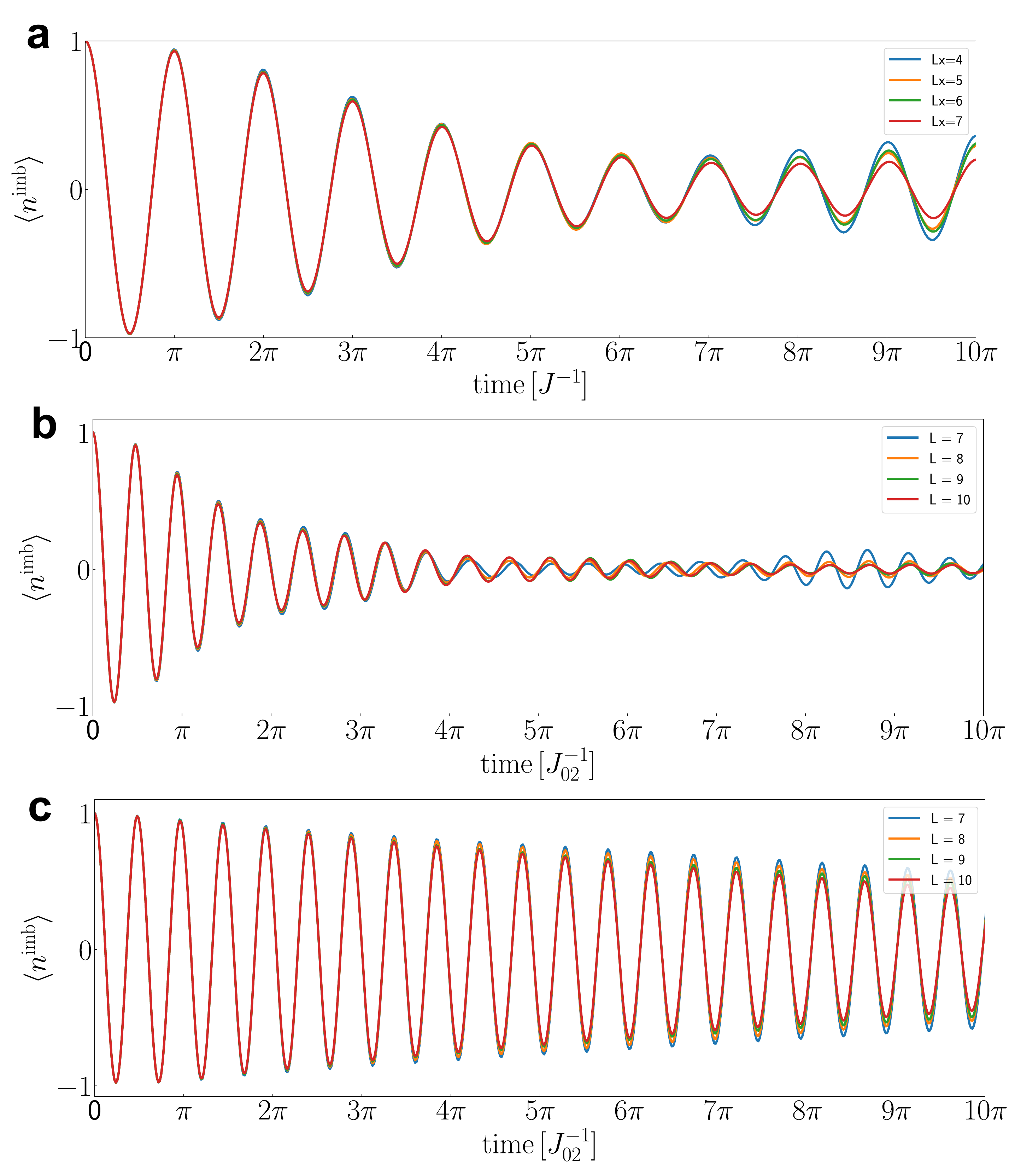}
    \caption{\label{fig:Sfinite}\textbf{Finite size effect on time evolution in Fig.~\ref{fig:2} and Fig.~\ref{fig:3} of the main text.} Here we show the imbalance dynamics of different system sizes under the same Hamiltonian. \textbf{(a)} Time evolution of imbalance under $U/J = U/J' = 50$ with bosonic lattice Hamiltonian. \textbf{(b)(c)} Time evolution of imbalance with the same configuration as in Fig.~\ref{fig:3}c without and with Floquet driving. } 
\end{figure}

For the time evolution results, we found that the initial time evolution is largely independent of system size, while for later time, the evolution is more sensitive. In order to test whether the results can be extended to even larger system, we compute the time evolution of exactly same parameters under different system sizes [see Fig.~\ref{fig:Sfinite}].  If the results are mostly the same for different system sizes, we believe it is a property that can be generalized to infinite size chain. 

\section{Optimizing QMBS in dipolar spin chains}
For a zigzag chain system, there are 5 degrees of freedom: the distance between site 0 and 1, $\alpha, \beta$, and the polar and azimuthal angles of the quantization axis. For the quantum many-body scar, there are two constraints: $J_{03} = 0$ and $J_{02}+J_{13} = 0$. Also, we want to fix the ratio of $J_{02}/J_{01}$ so that the model is completely determined. The distance between site 0 and 1 corresponds to an overall scaling of the interaction strength, which does not change the dynamical properties of the system. In the end, there is one degree of freedom for us to optimize, which we choose to be $\phi$, where $\phi$ is the angle between quantization axis and the plane. Here we demonstrate the optimization for $J_{02}/J_{01} = 0.3$. Results are shown in Fig.~\ref{fig:S2}. As $\phi$ increases, the long range term $J_{14}, J_{05}$ decreases, but at the same time, the absolute value of the interaction decreases (set $r_{01} = 1$, where $r_{01}$ is the distance between site 0 and 1). Ideally, we can keep increasing $\phi$ to reach a minimum strength of long range interactions. However, in a realistic experiment, we need to balance between minimizing long range interactions and keeping sufficiently high absolute interaction strengths.

\section{Leakage of QMBS due to experimental limitations and errors\label{sec:leakage}}
\subsubsection{Optical Lattice}
\begin{figure}[htb]
    \centering
    \includegraphics[width=0.6\linewidth]{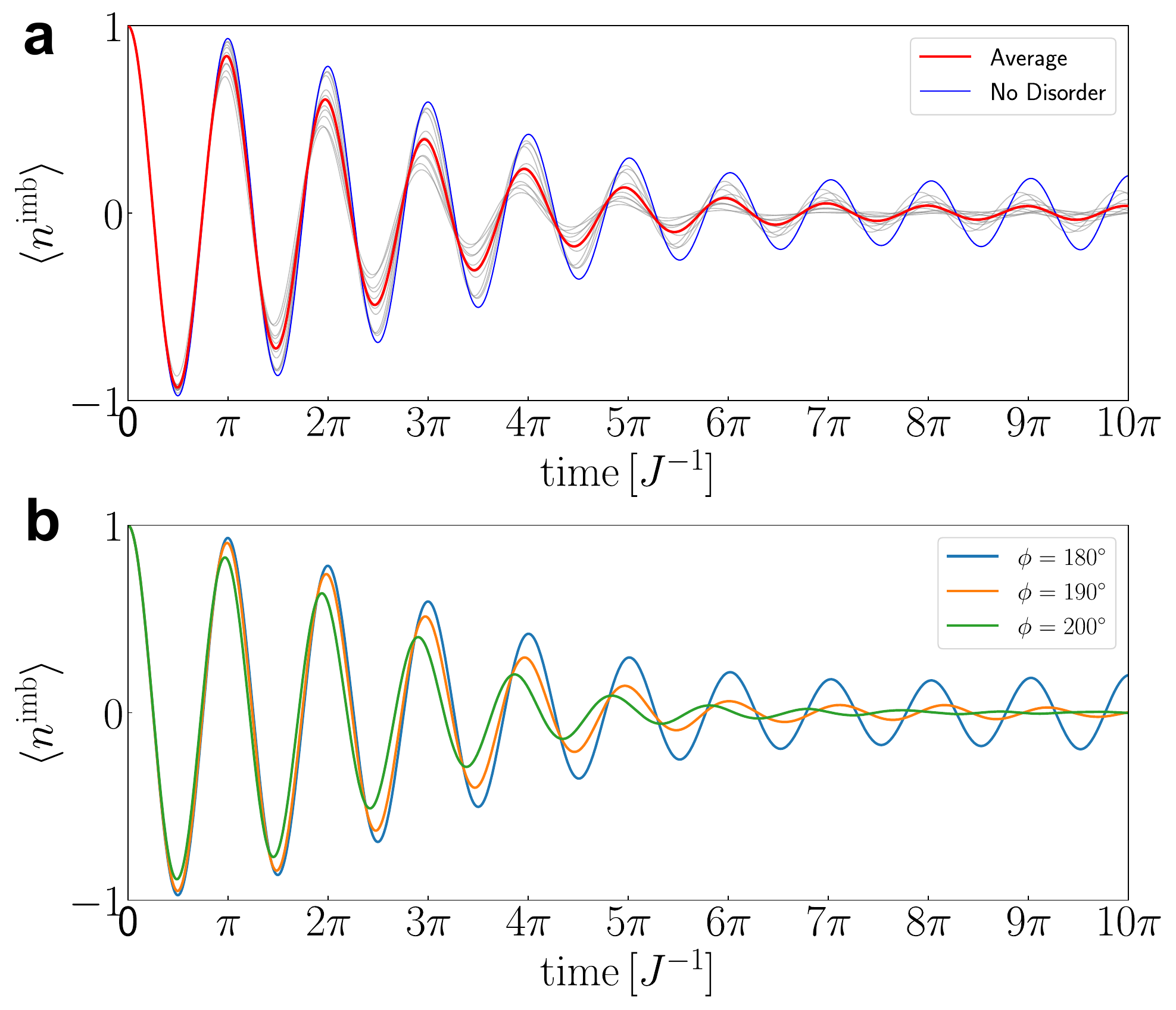}
    \caption{\textbf{(a)} Scarring dynamics for hardcore bosons on a $\pi$-flux ladder, with disorder in the chemical potential. The red curve shows the ideal case, where disorder is set to zero. The grey lines show the dynamics with a Gaussian distribution of chemical potentials [see text]. \textbf{(b)} Scarring dynamics with deviations to the $\phi=\pi$ phase for hardcore bosons in an optical $\pi$-flux lattice. The legend indicates the phase.}.
    \label{fig:S4}
\end{figure}

For state-of-the-art optical lattice experiments, possible errors include flux inhomogeneity and disorder in the chemical potential. We estimate the effects of these disorders in Fig.~\ref{fig:S4}.

\textit{Disorder in the chemical potential}: We assume a Gaussian distribution of chemical potentials in the lattice sites, with $\sigma = J/5$ and take $U/J = 50$ for all simulations. The simulation is carried out in a $L = 7 $ ladder.

\textit{Flux deviation:}
We assume the flux is uniform accross the ladder but has an overall deviation from $\pi$. The simulation is carried out with $U/J = 50$ in a $L = 7$ ladder.

\subsubsection{Rydberg atoms array}
\begin{figure}[htb]
    \centering
    \includegraphics[width=0.6\linewidth]{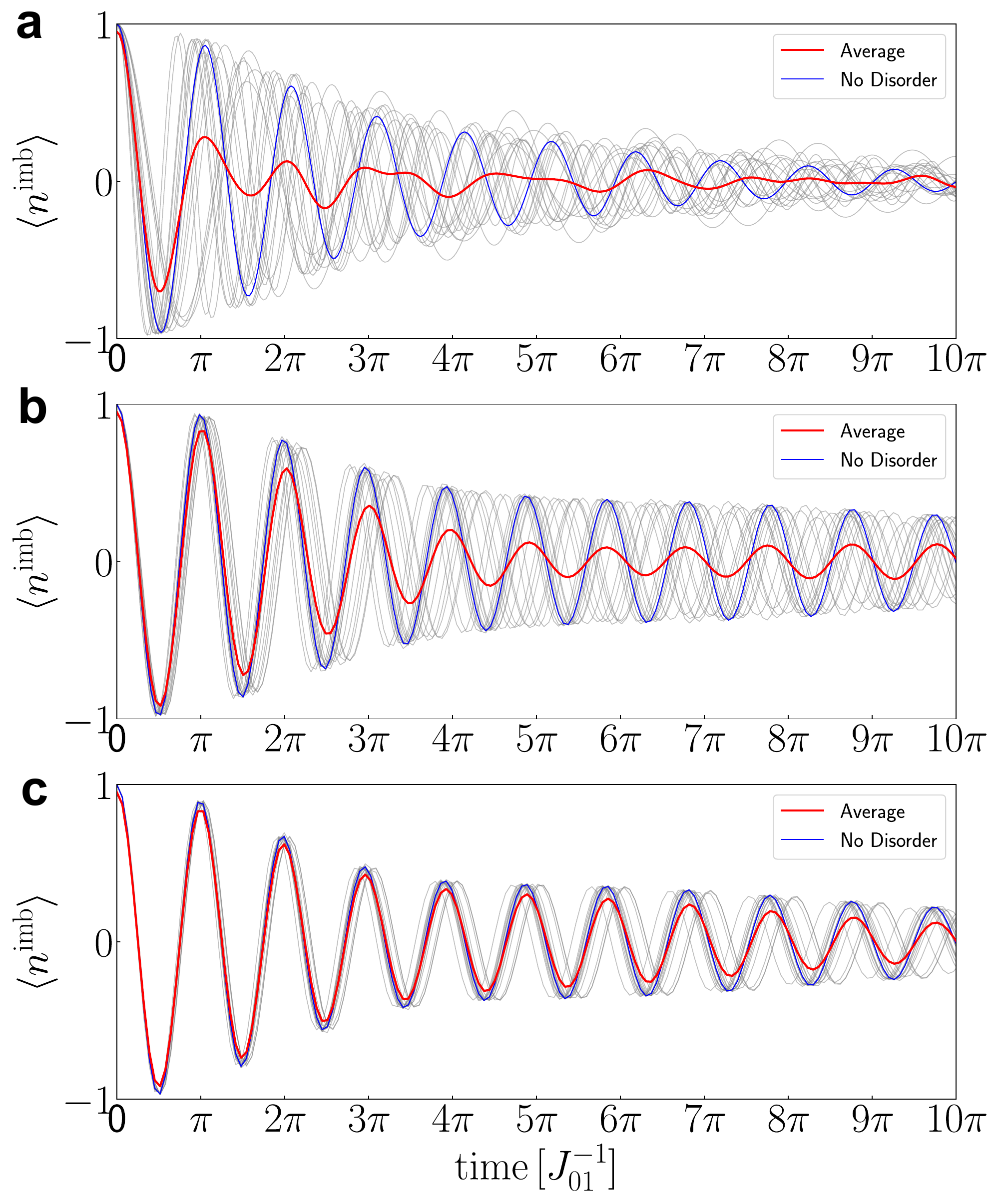}
    \caption{\textbf{Scarring dynamics for dipolar spin chain in the presence of positional disorder.} The gray curves show the results for individual runs of the experiment, \textit{i.e.}, different instances of random positional disorder. The red curves show the time evolution results averaged over 19 shots, and the blue curve shows the ideal result without disorder.  \textbf{(a)(b)(c)} correspond to different configurations. \textbf{(a)} $\phi = 30^\circ, \alpha = 34^\circ, J_{02}/J_{01} = 0.480$. \textbf{(b)} $\phi = 15^\circ, \alpha = 44^\circ, J_{02}/J_{01} = 0.275$. \textbf{(c)} $\phi = 0^\circ, \alpha = 52.49^\circ, J_{02}/J_{01} = 0.248$ .}
    \label{fig:S5}
\end{figure}

A dominant error in dipolar tweezer array experiments is positional disorder, whose effect is magnified when tweezers are moved closer to each other. We estimate the wavepacket spread of a single particle to be $\sigma_r = 100$\,nm, $\sigma_z = 800$\,nm, in the radial (in-plane) and axial (out-of-plane) directions respectively, and we set the distance between site 0 and 1 to 15$\mu$m. We simulate the results with 3 sets of parameters in a $L = 10$ ladder. Fig.~\ref{fig:S5} shows the imbalance dynamics averaged over multiple shots in the presence of positional disorder. 
We also note that for dipolar interactions $C_3/r^3$, further spacing the tweezers apart at the expense of reducing all $J$ couplings is one route to suppressing the effect of disorder. 

\section{Fidelity and Imbalance}
Here we would like to show the analytical expression of these two quantities and how to understand them. 
For a $L$ site chain, there are $L+1$ exact scar eigenstates, labeled as $\ket{\psi_n}$. The QMBS initial state can be written as $\ket{\psi_{\rm QMBS}} = \sum_n a_n\ket{\psi_n}$.
With static imperfections like long range interactions, $\ket{\psi_n}$ splits into several other states. Label the new eigenstates $\ket{\psi_{n\alpha}}$ and write $\ket{\psi_n} = \sum_\alpha b_{n\alpha} \ket{\psi_{n\alpha}}$. 
Define $\rho_{n\alpha}\equiv |b_{n\alpha}|^2$. The fidelity is simply given by
\begin{equation}
    F(t) = \left|\sum_{\alpha, n}a_n^2\rho_{n\alpha}e^{-iE_{n\alpha}t}\right|^2
\end{equation}
For the imbalance, notice that $\langle n^{\textrm{imb}} \rangle = \frac{1}{L}(J^+ + J^-)$. The averaged value of imbalance is given by
\begin{equation}
    \langle n^{\textrm{imb}} \rangle(t) =  \frac{1}{L2^{L-1}}\sum_{n = 0}^{L-1}\sum_{\alpha,\beta} \frac{L!}{n!(L-n-1)!}\rho_{n\alpha}\rho_{n+1,\beta}\cos(E_{n\alpha}-E_{n+1,\beta})t
\end{equation}

\section{Scar lifetime with perturbation strength}

\begin{figure}[htbp]
    \centering
    \includegraphics[width=0.9\linewidth]{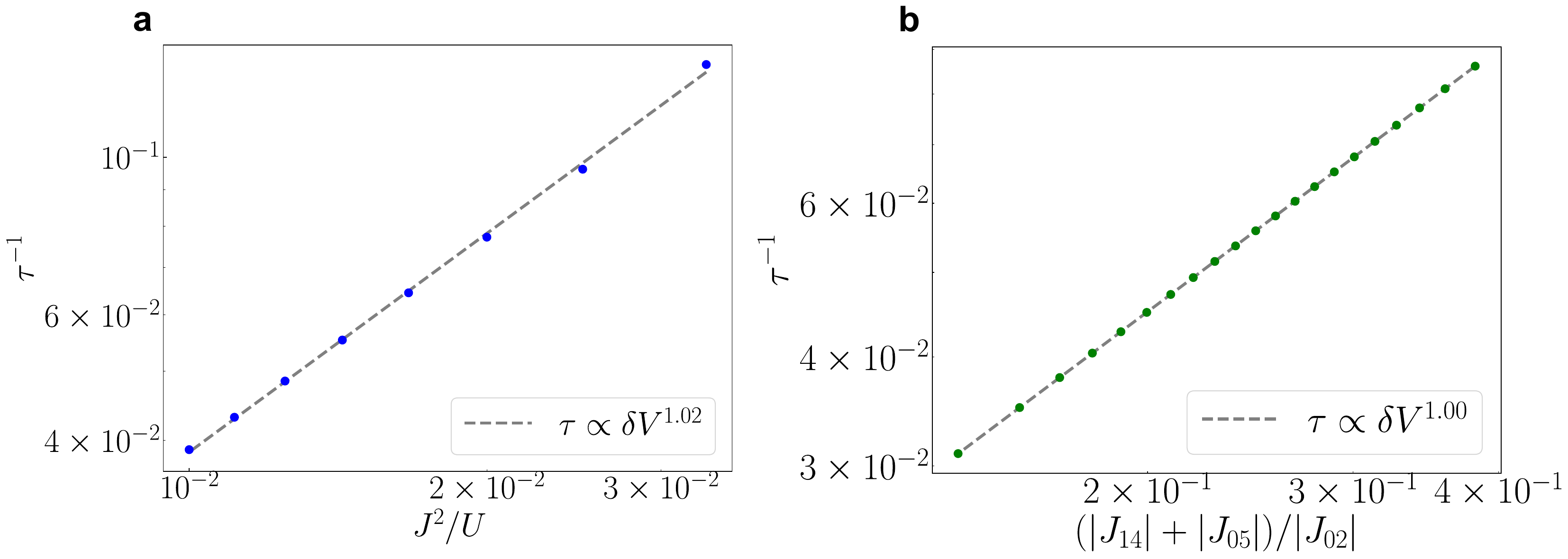}
    \caption{\label{fig:S7}\textbf{Scar lifetime with perturbation strength} \textbf{(a)} Inverse lifetime $\tau^{-1}$ with $\delta V = J^2/U$. An exponent of $1.022\pm 0.013$ was derived from logarithmic fitting. \textbf{(b)} Inverse lifetime $\tau^{-1}$ with $\delta V$. The exponent is $1.0013\pm 0.0001$.}
    \label{fig:S6}
\end{figure}

Here we investigate the relationship between scar lifetime and perturbation strength. The lifetime is estimated using the energy width $\sigma$ introduced in the main text. For bosonic lattice, the first order perturbation is {\tiny $\left| \begin{matrix} 0 & 0 \\ 1 & 1 \end{matrix} \right\rangle \to \left| \begin{matrix} 0 & 0 \\ 0 & 2 \end{matrix} \right\rangle \to \left| \begin{matrix} 0 & 0 \\ 1 & 1 \end{matrix} \right\rangle$}, with the perturbation strength $\delta V \propto J^2/U$. For dipolar spin chain, the perturbation terms are longer range interactions, with strength $\delta V \propto (|J_{14}|+|J_{05}|)/|J_{02}|$. The parameters for dipolar spin chain are from Fig~\ref{fig:S1}. We found that the lifetime $\tau$ proportional to $1/\delta V$ in both cases in Fig.~\ref{fig:S7}.

\end{document}